\pdfoutput=1
\documentclass[%
 reprint,
superscriptaddress,
 amsmath,amssymb,
 aps,
floatfix,
]{revtex4-2}

\newcommand{\angstrom}{\mbox{\normalfont\AA}}

\usepackage{braket}
\usepackage{listings}
\usepackage[caption=false]{subfig}  
\usepackage{graphicx}
\usepackage{dcolumn}
\usepackage{bm}
\usepackage{hyperref}
\usepackage{xcolor}

\begin{document}

\title{Analog quantum simulation of non-Condon effects in molecular spectroscopy}
\author{Hamza Jnane}
\affiliation{Télécom Paris, LTCI, 19 Place Marguerite Perey, 91120 Palaiseau, France}
\author{Nicolas P. D. Sawaya}
\affiliation{Intel Laboratories, Santa Clara, California 95054, United States}
\author{Borja Peropadre}
\affiliation{Zapata Computing Inc, 100 Federal St, Boston MA, 02110 USA.}
\author{Alan Aspuru-Guzik}
\affiliation{Department of Computer Science, University of Toronto, Toronto, Ontario M5S 2E4, Canada}
\affiliation{Department of Chemistry, University of Toronto, Toronto, Ontario M5G 1Z8, Canada}
\affiliation{Vector Institute for Artificial Intelligence, Toronto, Ontario M5S 1M1, Canada}
\affiliation{Canadian Institute for Advanced Research, Toronto, Ontario M5G 1Z8, Canada}
\author{Raul Garcia-Patron}
\affiliation{School of Informatics, University of Edinburgh, Edinburgh EH8 9AB, United Kingdom}
\author{Joonsuk Huh}
\email{joonsukhuh@gmail.com}
\affiliation{Department of Chemistry, Sungkyunkwan University, Suwon 16419, Republic of Korea}
\affiliation{SKKU Advanced Institute of Nanotechnology, Sungkyunkwan University, Suwon
16419, Republic of Korea}
\affiliation{Institute of Quantum Biophysics, Sungkyunkwan University, Suwon
16419, Republic of Korea}





\date{\today}

\begin{abstract}
In this work, we present a linear optical implementation for analog quantum simulation of molecular vibronic spectra, incorporating the non-Condon scattering operation with a quadratically small truncation error. 
Thus far, analog and digital quantum algorithms for achieving quantum speedup have been suggested only in the Condon regime, which refers to a transition dipole moment that is independent of nuclear coordinates. For analog quantum optical simulation beyond the Condon regime (i.e., non-Condon transitions) the resulting non-unitary scattering operations must be handled appropriately in a linear optical network. In this paper, we consider the first and second-order Herzberg-Teller expansions of the transition dipole moment operator for the non-Condon effect, for implementation on linear optical quantum hardware. We believe the method opens a new way to approximate arbitrary non-unitary operations in analog and digital quantum simulations. 
We report in-silico simulations of the vibronic spectra for naphthalene, phenanthrene, and benzene to support our findings.

\end{abstract}

\pacs{Valid PACS appear here}
\maketitle


\section{Introduction}
Quantum computing has attracted a considerable amount of attention over the past several decades due to its potential to accelerate research and development in fields such as  chemistry, cryptography, finance, and machine learning  \cite{Peev_2009,egger2020,Biamonte_2017}, by leveraging quantum mechanical properties, such as superposition and entanglement. 
Among these applications, chemistry is particularly important~\cite{cao2018,McArdle2020}, as it can impact other fields, such as biology, pharmacology, and material science through the development and increased understanding of molecular structures. For instance, quantum computers can potentially solve inextricable problems in biology ~\cite{Reiher2016}, and materials science~\cite{babbush2018,bauer2020}. Particularly, the computation of molecular vibronic spectra~\cite{jankowiak:2007,santoro:2007b,Huh2011a}, which has no known efficient classical algorithms, would be possible with the help of quantum machines ~\cite{huh2015,Huh2016VBS,sawaya2019,mcardle2019}. 
Simulating molecular vibronic spectra has a long history dating back to the beginning of quantum chemistry.
Vibronic spectroscopy is important as it can provide insightful information about a molecule's optical properties, which are essential to biological applications \cite{butler2016} and solar-cells \cite{hachmann2011}, among other applications. The optoelectronic molecular process involves transitions between two different Born-Oppenheimer (BO) electronic potential energy surfaces (PESs), each of which embeds vibrational manifolds. The transition probabilities are called Franck-Condon factors, which are the square modulus of the overlap integral between two vibrational wavefunctions belonging to the two different PESs. The quantities can be obtained as multivariate normal moments or multivariate Hermite polynomials or hafnians within the harmonic approximation; however, the best-known algorithms scale exponentially with the size of the problem~\cite{kan:2007,Huh2011a,huh2020b,Quesada2019}. 



Analog~\cite{huh2015,Huh2016VBS} and digital~\cite{sawaya2019,mcardle2019,sawaya2020_dlev} quantum algorithms have been proposed to simulate molecular vibronic spectra. Although the quantum phase estimation algorithm-based digital approach~\cite{mcardle2019,sawaya2019} considers anharmonicity (which is difficult to simulate classically \cite{huh2010,luis2006,Meier2015,petrenko2017}), it requires a fault-tolerant quantum computer to output satisfactory results. Although it may be possible to create another digital quantum protocol, that is suitable for noisy intermediate-scale quantum devices,  such as the variational quantum eigensolver~\cite{mcardle2019, ollitrault2020,sawaya2020_spec}, the non-Condon problem may still require significant resources. In contrast, an analog quantum simulator would be experimentally implementable for molecules of moderate size~\cite{Peropadre2015,Shen2017,Clements2018,wang2019} to bypass the technological obstacles of building a universal quantum computer. The analog quantum simulator was initially proposed in the quantum optical language (i.e., Gaussian operators) to consider linear optical sampling problems, such as boson sampling \cite{Aaronson2011}. 

Huh et al.~\cite{huh2015} proposed that a Gaussian boson sampling~\cite{rahimi2015,kruse2019} setup can naturally simulate a molecular vibronic spectrum assuming a constant transition dipole moment (TDM), also known as Condon approximation, and harmonic approximation of BO surfaces.  Although the complexity of the molecular problem is related to the Gaussian boson sampling problem, yet it remains unclear \cite{cao2018}. The Gaussian boson sampling setup has been implemented~\cite{huh2020} for small molecules in multiple quantum devices such as trapped ions \cite{Shen2017}, linear optics \cite{Clements2018}, and superconducting circuits \cite{wang2019}. 

Although the proposal by Huh et al.~\cite{huh2015} has a quantum advantage and practical applications, classical simulation can be achieved in practice for hundred-atom molecules \cite{jankowiak:2007,santoro:2007b} within the harmonic picture, which is far beyond the realizable size of quantum devices, using clever strategies for cutting-off required integral calculations. Therefore, to be useful to chemists, a quantum simulator must demonstrate more complex molecular processes. 
The fundamental goal of quantum simulation, which is quantum speedup with practical applications, should be approached by releasing existing restrictions, namely, the Condon and harmonic approximations. In this work, we focus on the non-Condon effect, as a first challenge, which invokes the coordinate dependence of the TDM. To incorporate the non-Condon effect, we start from the Taylor expansion  \cite{Herzberg1933,small1971,baiardi:2013,santoro:2007b,baiardi:2013} of the TDM operator with respect to the normal coordinates of molecules (i.e., Herzberg-Teller (HT) expansion). In this setup, the probabilities for each transition require an even larger number of integral calculations than the Condon case: approximately 3-7 times more Franck-Condon integral calculations are required for the first and second-order HT integrals. Here, we propose a Gaussian boson sampling approach to simulate the non-Condon spectrum for the linear and quadratic HT terms. The main challenge is determining how to handle the non-unitary operation (non-Condon operator), which is non-trivial in a linear optics quantum simulator using only passive elements such as beam splitters and phase shifters. To overcome this problem, we construct the non-Condon profile as a linear combination of four independent sampling problems (see Fig.1 for an illustration) for each polarization direction of TDM.

The remainder of this paper is organized as follows. In Section II, we describe the non-Condon vibronic transition. Then, in Section III, we present how the computation of the non-Condon profile can be transformed into a quantum optical sampling problem. 
Finally, in Section IV, we demonstrate numerically obtained vibronic spectra of  naphthalene, phenanthrene, and benzene.

\section{\label{sec:Section vib} Non-Condon profile}

When a molecule absorbs a photon, the molecule undergoes simultaneously vibrational and electronic state changes, also called vibronic transitions. Within the BO approximation (separation of electronic and nuclear degrees of freedom), determining the vibronic spectral profile involves computing the transition probabilities between any two vibrational states of the initial and excited PESs. The PESs are generally anharmonic; however, in this work, we approximate them as harmonic PESs. 

Because the vibronic transition profile at finite temperature can always be transformed into the zero-temperature profile by purifying the initial thermal state~\cite{Huh_2017}, we focus on the zero-temperature non-Condon transition profile for simplicity, which is given as a Fermi's golden rule type equation: 
\begin{align}
P(\omega)=\frac{1}{\mathcal{N}}\sum_{\substack{\mathbf{m^{'}}=~\mathbf{0} \\r=x,y,z}}^{\boldsymbol{\infty}}\vert\langle\mathbf{m^{'}}\vert\hat{U}_{\mathrm{Dok}}\hat{\mu}_{r}(\hat{\mathbf{Q}})\vert\mathbf{0}\rangle\vert^{2}\delta(\omega-\mathbf{m^{'}}\cdot\boldsymbol{\omega}'),
\label{eq:vibronicprofile}
\end{align}
where $\mathcal{N} = \sum_{r =x,y,z}\mathcal{N}_{r}$ is a normalization constant, that depends on the TDM operator $\hat{\boldsymbol{\mu}}(\mathbf{\hat Q}) (=(\hat{\mu}_{x}(\hat{\mathbf{Q}}),\hat{\mu}_{y}(\hat{\mathbf{Q}}),\hat{\mu}_{z}(\hat{\mathbf{Q}}))^{\mathrm{t}})$ with respect to the mass-weighted normal coordinates of the initial state $\mathbf{\hat Q}=(\hat{Q}_{1},\ldots,\hat{Q}_{M})^{\mathrm{t}}$, where the corresponding frequency vector is $\boldsymbol{\omega}=(\omega_{1},\ldots,\omega_{M})^{\mathrm{t}}$.  $\vert\mathbf{m^{'}}\rangle=\vert m^{'}_{1},\ldots,m^{'}_{M}\rangle$ and $\vert\mathbf{0}\rangle=\vert 0,\ldots,0\rangle$ are the $M$-dimensional final Fock state and initial vacuum state, respectively. $\omega$ is the transition frequency satisfying the resonance condition $\omega=\mathbf{m^{'}}\cdot\boldsymbol{\omega}'$, where $\boldsymbol{\omega}'=(\omega_{1}',\ldots,\omega_{M}')^{\mathrm{t}}$ is the $M$-dimensional harmonic frequency vector of the normal coordinates ($\hat{\mathbf{Q}}'=(\hat{Q}_{1}',\ldots,\hat{Q}_{M}')^{\mathrm{t}}$) of the final electronic state. 
The normal coordinates of the final and initial electronic states are linearly related by the Duschinsky transformation~\cite{duschinsky:1937} $\hat{\mathbf{Q}}'=\mathbf{U_\mathrm{D}}\hat{\mathbf{Q}}+\mathbf{d}$, where $\mathbf{d}$ is a molecular geometric displacement vector. The Duschinsky transformation can be represented by the Doktorov operator $\hat{U}_{\mathrm{Dok}}$~\cite{doktorov:1977}, as a unitary transformation of the form $\mathbf{\hat{Q}^{'}}=\hat{U}_{\mathrm{Dok}}^{\dagger} \mathbf{\hat{Q}}\hat{U}_{\mathrm{Dok}} $. We can rewrite the Duschinsky transformation as a multidimensional Bogoliubov transformation as follows:
\begin{align}
    \hat{\mathbf{a}}^{'\dagger} = \frac{1}{2}\left(\mathbf{J}-(\mathbf{J}^{\mathrm{t}})^{-1}\right)\hat{\mathbf{a}} + \frac{1}{2}\left(\mathbf{J}+(\mathbf{J}^{\mathrm{t}})^{-1}\right)\hat{\mathbf{a}}^{\dagger}+\frac{1}{\sqrt{2}}\boldsymbol{\delta}
\end{align}
where $\mathbf{\hat a} = (\hat a_1,\ldots,\hat a_M)^\mathrm{t}$, and $\hat{a}_{j}$ and $\hat{a}_{j}^{\dagger}$ are the boson annihilation and creation operators, respectively, satisfying the boson commutation relation $[\hat{a}_{i},\hat{a}_{j}^{\dagger}]=\delta_{ij}$. In addition, $\mathbf{J} = \mathbf{\Omega^{'}}\mathbf{U_\mathrm{D}}\mathbf{\Omega}^{-1},
\mathbf{\Omega^{'}} = \mathrm{diag}(\sqrt{\omega_1^{'}},\cdots,\sqrt{\omega_M^{'}}), \mathbf{\Omega} = \mathrm{diag}(\sqrt{\omega_1},\cdots,\sqrt{\omega_M})$ and $\boldsymbol{\delta} = \hbar^{-\tfrac{1}{2}}\mathbf{\Omega^{'}}\mathbf{d}$.

Eq.~\eqref{eq:vibronicprofile} tells that the profile can be written as three independent sums, one for each polarization direction. For the sake of simplicity, we consider only a single component of the TDM operator $\hat{\mu}_r$ ($r \in \{x,y,z\}$) and thus a single sum. In the following, we focus on the $x$-direction and drop the subscript.

The TDM operator is generally dependent on the nuclear coordinate of the molecule. To account for the coordinate dependence, we expand the TDM with respect to $\hat{\mathbf{Q}}$, also known as HT expansion.
Usually, the zeroth-order term is the most significant, and the coordinate dependence is often ignored in Franck-Condon approximation. However, in some cases, the vibronic spectra exhibit significant coordinate dependence of the TDM. Thus, higher-order terms must be considered for these transitions. 

For some transitions, the zeroth-order term $\mu^{(0)}$ may be close to or equal to zero. These are called Franck-Condon forbidden transitions ($\vert\mu^{(0)}\vert=0$) or weakly-allowed Franck-Condon transitions ($\vert\mu^{(0)}\vert\ll 1$). 
For example~\cite{Dierksen2004}, the vibronic transitions of anthracene ($1~^{1}\mathrm{A_{g}}$--$1~^{1}\mathrm{B_{2u}}$), pentacene ($1~^{1}\mathrm{A_{g}}$--$1~^{1}\mathrm{B_{2u}}$), pyrene ($1~^{1}\mathrm{A_{g}}$--$1~^{1}\mathrm{B_{2u}}$), octatetracene ($1~^{1}\mathrm{A_{g}}$--$1~^{1}\mathrm{B_{u}}$), and styrene ($1~^{1}\mathrm{A'}$--$3~^{1}\mathrm{A'}$) are strongly dipole allowed transitions (Franck-Condon allowed, $\vert\mu^{(0)}\vert\simeq 1$);  
pyrene ($1~^{1}\mathrm{A_{g}}$--$1~^{1}\mathrm{B_{3u}}$),  
azulene ($1~^{1}\mathrm{A_{1}}$--$1~^{1}\mathrm{B_{1}}$), and  
phenoxyl radical ($1~^{2}\mathrm{B_{1}}$--$1~^{2}\mathrm{A_{2}}$) are weakly dipole-allowed transitions; octatetraene ($1~^{1}\mathrm{A'}$--$2~^{1}\mathrm{A'}$), and benzene ($^{1}\mathrm{B_{2u}}$--$^{1}\mathrm{A_{1g}}$) 
are dipole-forbidden transitions. 
When $\hat{\mu}(\hat{\mathbf{Q}})$ is expanded up to the linear term, the first-order HT expansion is obtained as follows:
\begin{align}
\hat{\mu}(\hat{\mathbf{Q}})&=\mu^{(0)}
+\sum_{j=1}^{M}\mu_{j}^{(1)}\hat{Q}_{j}+\cdots \nonumber \\
&\simeq \mu^{(0)}
+\boldsymbol{\lambda}^{\mathrm{t}}\cdot\frac{(\hat{\mathbf{a}}+\hat{\mathbf{a}}^{\dagger})}{\sqrt{2}} ,
\end{align}
where $\lambda_{j}=\sqrt{\tfrac{\hbar}{\omega_{j}}}\mu_{j}^{(1)}$
is due to the relation $\hat{Q}_{j}=\sqrt{\tfrac{\hbar}{2\omega_{j}}}(\hat{a}_{j}+\hat{a}_{j}^{\dagger})$. For the second-order HT expansion,
\begin{align}
&\hat{\mu}(\hat{\mathbf{Q}}) = \mu^{(0)}
+\sum_{j=1}^{M}\mu_{j}^{(1)}\hat{Q}_{j}+\frac{1}{2}\sum_{j,k=1}^{M}\mu_{j,k}^{(2)}\hat{Q}_{j}\hat{Q}_{k} + \cdots \nonumber \\
&\simeq \mu^{(0)}
+\boldsymbol{\lambda}^{\mathrm{t}}\cdot\frac{(\hat{\mathbf{a}}+\hat{\mathbf{a}}^{\dagger})}{\sqrt{2}}
+\left(\frac{\hat{\mathbf{a}}+\hat{\mathbf{a}}^{\dagger}}{\sqrt{2}}\right)^{\mathrm{t}}
\boldsymbol{\Lambda}\left(\frac{\hat{\mathbf{a}}+\hat{\mathbf{a}}^{\dagger}}{\sqrt{2}}\right) ,
\end{align}
where $[\boldsymbol{\Lambda}]_{j,k}=\tfrac{\mu_{j,k}^{(2)}}{2}\sqrt{\tfrac{\hbar}{\omega_{j}}}\sqrt{\tfrac{\hbar}{\omega_{k}}}$ and $\boldsymbol{\Lambda}=\boldsymbol{\Lambda}^{\mathrm{t}}$. The required molecular parameters can be obtained from quantum chemistry calculations~\cite{berger:1998,santoro:2008}. 
The normalization factors (for the $x$-direction only) of the first-order and second order expansions are given by $\mathcal{N}_1 = (\mu^{(0)})^2 + \tfrac{1}{2}\sum_{j=1}^M\lambda_j^2$ and $\mathcal{N}_2 = (\mu^{(0)})^2 + \tfrac{1}{2}\sum_{j=1}^M\lambda_j^2 +\mu^{(0)}\sum_{j=1}^{M}[\boldsymbol{\Lambda}]_{j,j}+ \tfrac{1}{4}\sum_{\substack{j,k=1\\ j \neq  k}}^{M}[\boldsymbol{\Lambda}]_{j,j}[\boldsymbol{\Lambda}]_{k,k}+\tfrac{1}{2}\sum_{\substack{j,k=1\\ j \neq  k}}^{M} [\boldsymbol{\Lambda}]_{j,k}^2 + \tfrac{3}{4}\sum_{j=1}^{M} [\boldsymbol{\Lambda}]_{j,j}^2 $, respectively (see Appendix \ref{app:appendix Norm_const} for the derivation).

\section{\label{sec:Section approx}Approximating non-Condon profile with Gaussian states}
To use a linear quantum optics simulator for the non-Condon profile, we must implement  $\hat{U}_{\mathrm{Dok}}\hat{\mu}(\hat{\mathbf{Q}})$ in the linear optical setup and measure the output photon states $\vert\mathbf{m^{'}}\rangle$. However, we cannot achieve this directly with the Gaussian boson sampling device as $\hat{\mu}(\hat{\mathbf{Q}})$ is a non-unitary and non-Gaussian operator. Here, we use the following multimode-operator notation \cite{Ma1990} for the displacement operator $\hat D$,  squeezing operator $\hat S$, and  rotation operator $\hat R$: 
\begin{align}
    \hat D(\boldsymbol \alpha) &= \exp \left(\boldsymbol \alpha^\mathrm{t} \hat{\mathbf a}^{\dagger}-\boldsymbol{\alpha}^\dagger \hat{\mathbf a}\right), \\ 
    \hat S(\boldsymbol \Xi) &= \exp \left(\frac{1}{2}((\mathbf{\hat a}^\dagger)^{\mathrm{t}}\boldsymbol \Xi\hat{\mathbf a}^{\dagger}-\hat{\mathbf a}^{\mathrm{t}} \boldsymbol{\Xi}^\dagger \hat{\mathbf a})\right), \\
    \hat R(\mathbf{U}) &= \exp\left((\mathbf{\hat a}^\dagger)^{\mathrm{t}}\ln(\mathbf{U}^*)\mathbf{\hat a}\right),
\end{align}
where $\boldsymbol \alpha$, $\boldsymbol \Xi$ are a complex vector and complex matrix, respectively, and $\mathbf{U}$ is a unitary matrix. 
Their action on the ladder operators is given by \cite{Ma1990},
\begin{align}
    \hat S(\boldsymbol{\Xi})^{\dagger}\hat {\mathbf{a}}\hat S(\boldsymbol{\Xi}) &= \mathrm{cosh}(\boldsymbol{\Xi})\hat{\mathbf{a}} + \mathrm{sinh}(\boldsymbol{\Xi})\hat{\mathbf{a}}^{\dagger}, \\
    \hat D(\boldsymbol{\alpha})^{\dagger} \hat{\mathbf{a}} \hat D(\boldsymbol{\alpha}) &= \hat{\mathbf{a}} + \boldsymbol{\alpha}, \\ 
    \hat{R}(\mathbf{U})^{\dagger}\hat{\mathbf{a}}\hat{R}(\mathbf{U})&=\mathbf{U}\hat{\mathbf{a}},
\end{align}
where the shorthand notation $\hat O^\dagger \mathbf{\hat a} \hat O = (\hat O^\dagger \hat{a}_1 \hat O,\ldots,\hat O^\dagger \hat{a}_M \hat O)^{\mathrm{t}}$ is used for any operator $\hat O$.

In the following subsection, we introduce an auxiliary function to incorporate the non-Condon operator in a linear optical network. The auxiliary function is Taylor-expanded such that the linear combination of the auxiliary functions results in an approximation of the non-Condon transition with a quadratically scaling in parameter kappa. In this way, we can embed our problem into a Gaussian boson sampling device. 

\subsection{Gaussian approximation of  non-Condon transition}
We introduce an auxiliary function,
\begin{align}
f_{\mathbf{m^{'}}}(\kappa) = \vert\langle \mathbf{m^{'}}\vert\hat{U}_{\mathrm{Dok}}\exp(\kappa\hat{\mu})\vert\mathbf{0}\rangle\vert^{2},
\end{align}
where $\hat{\mu}=\hat{\mu}^{\dagger}$
and $\kappa\in\mathbb{C}$, to approximate the non-Condon profile in the form of a Gaussian boson sampler~\cite{huh2015}. For simplicity, the coordinate dependence of $\hat{\mu}$ is not indicated in this subsection. By expanding $f_{\mathbf{m^{'}}}$ up to the fourth order with respect to $\kappa$, we can approximate  $\vert\langle\mathbf{m^{'}}\vert\hat{U}_{\mathrm{Dok}}\hat{\mu}\vert\mathbf{0}\rangle\vert^{2}$ as follows: 
\begin{align}
f_{\mathbf{m^{'}}}(\kappa)&=\vert\langle \mathbf{m^{'}}\vert \hat{U}_{\mathrm{Dok}}\vert\mathbf{0}\rangle\vert^{2}\nonumber\\
&+2\mathrm{Re}(\kappa)\langle \mathbf{m^{'}}\vert \hat{U}_{\mathrm{Dok}}\hat{\mu}\vert\mathbf{0}\rangle
\langle \mathbf{0}\vert \hat{U}_{\mathrm{Dok}}^{\dagger}\vert\mathbf{m^{'}}\rangle \nonumber\\
&+\mathrm{Re}(\kappa^{2})\langle \mathbf{m^{'}}\vert \hat{U}_{\mathrm{Dok}}\hat{\mu}^{2}\vert\mathbf{0}\rangle
\langle \mathbf{0}\vert \hat{U}_{\mathrm{Dok}}^{\dagger}\vert\mathbf{m^{'}}\rangle\nonumber\\
&+\vert\kappa\vert^{2}\langle \mathbf{m^{'}}\vert \hat{U}_{\mathrm{Dok}}\hat{\mu}\vert\mathbf{0}\rangle
\langle \mathbf{0}\vert \hat{\mu}\hat{U}_{\mathrm{Dok}}^{\dagger}\vert\mathbf{m^{'}}\rangle \nonumber\\
&+\frac{1}{3}\mathrm{Re}(\kappa^{3})\langle \mathbf{m^{'}}\vert \hat{U}_{\mathrm{Dok}}\hat{\mu}^{3}\vert\mathbf{0}\rangle
\langle \mathbf{0}\vert \hat{U}_{\mathrm{Dok}}^{\dagger}\vert\mathbf{m^{'}}\rangle \nonumber\\
&+\frac{1}{2}\mathrm{Re}(\kappa^{2}\kappa^{*})\langle \mathbf{m^{'}}\vert \hat{U}_{\mathrm{Dok}}\hat{\mu}^{2}\vert\mathbf{0}\rangle
\langle \mathbf{0}\vert \hat{\mu}\hat{U}_{\mathrm{Dok}}^{\dagger}\vert\mathbf{m^{'}}\rangle \nonumber\\
&+\frac{1}{12}\mathrm{Re}(\kappa^{4})\langle \mathbf{m^{'}}\vert \hat{U}_{\mathrm{Dok}}\hat{\mu}^{4}\vert\mathbf{0}\rangle
\langle \mathbf{0}\vert \hat{U}_{\mathrm{Dok}}^{\dagger}\vert\mathbf{m^{'}}\rangle \nonumber\\
&+\frac{1}{3}\mathrm{Re}(\kappa^{3}\kappa^{*})\langle \mathbf{m^{'}}\vert \hat{U}_{\mathrm{Dok}}\hat{\mu}^{3}\vert\mathbf{0}\rangle
\langle \mathbf{0}\vert \hat{\mu}\hat{U}_{\mathrm{Dok}}^{\dagger}\vert\mathbf{m^{'}}\rangle \nonumber\\
&+\frac{1}{4}\vert\kappa\vert^{4}\langle \mathbf{m^{'}}\vert \hat{U}_{\mathrm{Dok}}\hat{\mu}^{2}\vert\mathbf{0}\rangle
\langle \mathbf{0}\vert \hat{\mu}^{2}\hat{U}_{\mathrm{Dok}}^{\dagger}\vert\mathbf{m^{'}}\rangle \nonumber\\
&+\mathcal{O}(\kappa^{5}).
\end{align}
Then, using a linear combination of $f_{\mathbf{m^{'}}}$ for different values of $\kappa$, we obtain
\begin{align}
    &\vert\langle \mathbf{m^{'}}\vert\hat{U}_{\mathrm{Dok}}\hat{\mu}\vert\mathbf{0}\rangle\vert^{2}= \nonumber \\
        &\frac{1}{2\tau^{2}}\left(f_{\mathbf{m^{'}}}(\mathrm{i}\tau)+\frac{1}{2}f_{\mathbf{m^{'}}}(\tau)+\frac{1}{2}f_{\mathbf{m^{'}}}(-\tau)-2f_{\mathbf{m^{'}}}(0)\right)+\mathcal{O}(\tau^{2})
    ,
\label{eq:linear_comb_aux_func}
\end{align}
where $\tau$ (given as the Debye inverse, $\mathrm{D}^{-1}$) is a real positive number resulting the quadratic truncation error $O(\tau^2)$ for approximating the non-unitary operator as a linear combination of unitary operators. Though this signifies that a smaller $\tau$ leads to smaller error, the smallest possible $\tau$ will be dictated by the precision allowed by a given device.

Now, the approximation of the non-Condon vibronic profile allowing the quadratic error requires a single evaluation of the Frank-Condon profile ($ f_{\mathbf{m^{'}}}(0)=\vert\langle \mathbf{m^{'}}\vert \hat{U}_{\mathrm{Dok}}\vert\mathbf{0}\rangle\vert^{2}$)~\cite{huh2015} and three evaluations of  $f_{\mathbf{m^{'}}}$, which can also be implemented as a Gaussian boson sampler~\cite{huh2015}. In the following section, we describe how $f_{\mathbf{m^{'}}}$ can be expressed with Gaussian operators to be implemented in linear optical networks.

\subsection{Non-Condon profile with  Gaussian boson sampler}
\begin{figure*}
    \centering
    \includegraphics[width=0.9\textwidth,height=0.9\textheight,keepaspectratio]{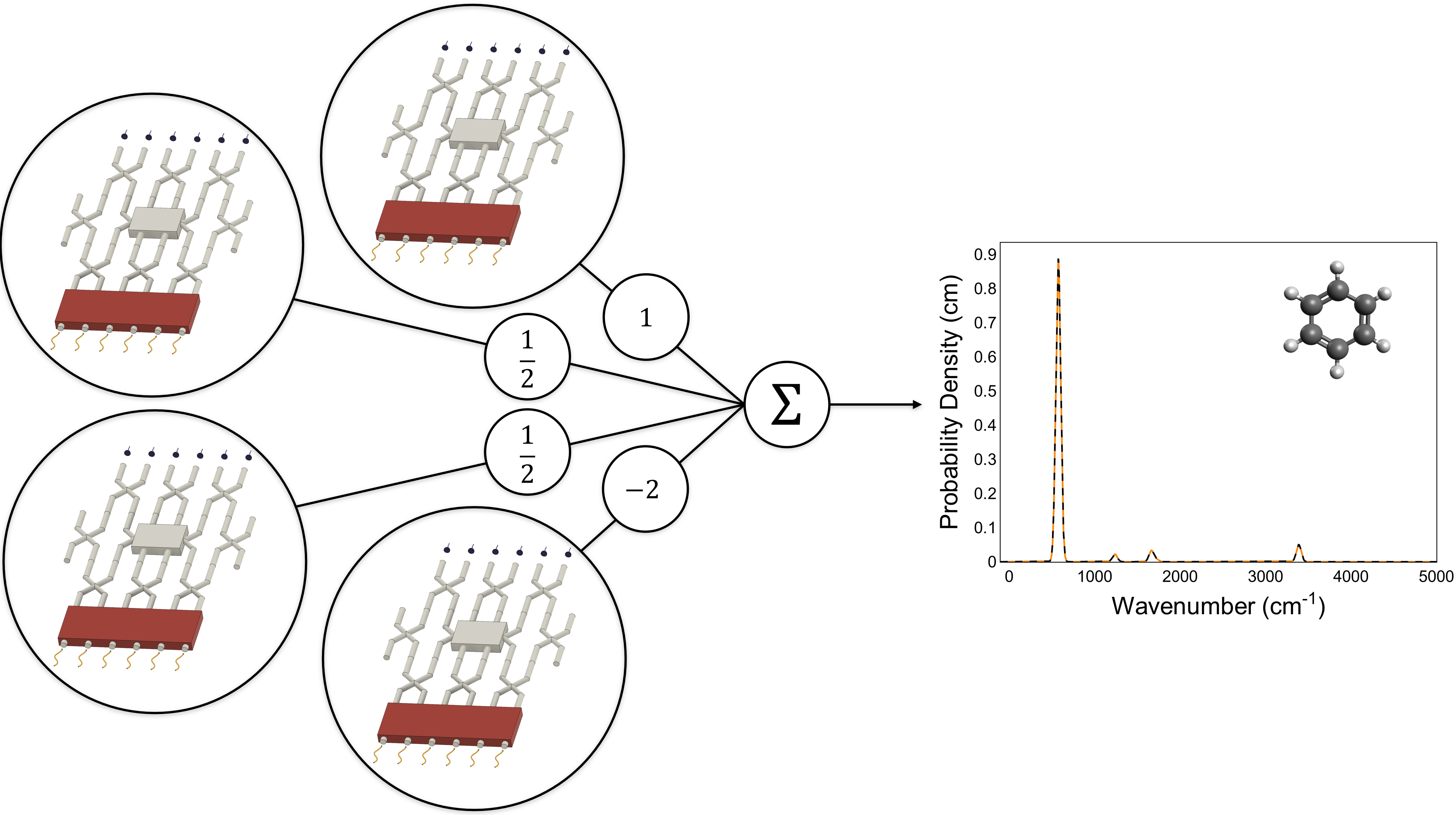}
    \caption{Illustration of a Gaussian boson sampler for simulating the non-Condon transition in Eq.~\eqref{eq:vibronic_profile_lin_comb}. The yellow wiggles represent vacuum states. By applying Gaussian operations, the squeezed coherent states are prepared before entering the interferometers (gray) characterized by unitary matrices. After collecting the output photon number distributions of the four independent linear optical networks using photon number resolving detectors (blue), we can reconstruct the non-Condon profile by a linear combination ($\Sigma$) of the measurement data with the weights in circles. The plotted spectrum is benzene's linear HT vibronic profile of the e2g symmetry block (see the numerical simulation section for details).
   }
    \label{fig:GBS_vibronic_profile}
\end{figure*}

We derive $f_{\mathbf{m^{'}}}$ for the second-order HT expansion, as the first-order case can be obtained simply by setting $\boldsymbol{\Lambda}=0$. Let $\hat{\mathbf{q}}=\tfrac{1}{\sqrt{2}}(\hat{\mathbf{a}}+\hat{\mathbf{a}}^{\dagger})$.  As $\boldsymbol{\Lambda}$ is symmetric, we have, $\boldsymbol{\Lambda}=\mathbf{U}^{\mathrm{t}}\mathbf{DU}$, where $\mathbf{U}$ is unitary. Expressing the TDM operator as a function of $\mathbf{\hat q}$ leads to, 
\begin{align}
\hat{\mu}(\hat{\mathbf{q}})&= \mu^{(0)}
+\boldsymbol{\lambda}^{\mathrm{t}}\cdot\hat{\mathbf{q}}
+\hat{\mathbf{q}}^{\mathrm{t}}
\boldsymbol{\Lambda}\hat{\mathbf{q}} \nonumber \\
&=  \hat{R}(\mathbf{U}^\mathrm{t})
\left[\mu^{(0)}
+\mathbf{b}^{\mathrm{t}}\cdot\hat{\mathbf{q}}
+\hat{\mathbf{q}}^{\mathrm{t}}\mathbf{D}\hat{\mathbf{q}}\right]\hat{R}(\mathbf{U}^\mathrm{t})^{\dagger} ,
\label{eq:q-expansion_rotated}
\end{align}
where $\mathbf{D}=\mathrm{diag}(d_{1},\ldots,d_{M})$, $\mathbf{b}=\mathbf{U}\boldsymbol{\lambda}$ and $\hat{R}(\mathbf{U}^\mathrm{t})\hat{\mathbf{a}}^{\dagger}\hat{R}(\mathbf{U}^\mathrm{t})^{\dagger}=\mathbf{U}\hat{\mathbf{a}}^{\dagger}$.
Writing $\hat{\mu}$ in this form allows us to easily express the action of its exponential on the vacuum state, i.e.
\begin{align}
&\exp(\kappa\hat{\mu})\vert\mathbf{0}\rangle \nonumber \\
&=\hat{R}(\mathbf{U}^\mathrm{t})\exp(\kappa(\mu^{(0)}
+\mathbf{b}^{\mathrm{t}}\cdot\hat{\mathbf{q}}
+\hat{\mathbf{q}}^{\mathrm{t}}\mathbf{D}\hat{\mathbf{q}}))\vert\mathbf{0}\rangle
\nonumber \\
&=\exp(\kappa\mu^{(0)})\hat{R}(\mathbf{U}^\mathrm{t})\bigotimes_{j=1}^{M}\exp(\kappa b_{j}\hat{q}_{j}+\kappa d_{j}\hat{q}_{j}^{2})\vert 0\rangle_{j} ,
\end{align}
where we use $\hat R(\mathbf{U}) \ket{\mathbf{0}} = \ket{\mathbf{0}}$.
For a single mode $j$, $\exp(\kappa b_{j}\hat{q}_{j}+\kappa d_{j}\hat{q}_{j}^{2})\vert 0\rangle_{j}$ can be decomposed  with Gaussian operators~\cite{Fan2003} as follows:  
\begin{align}
    \exp(\kappa b_{j}\hat{q}_{j}+\kappa d_{j}\hat{q}_{j}^{2})\ket{0}_j = C_{j,\kappa} \hat S_j(\xi_{j,\kappa})\hat D_j\left(\alpha_{j,\kappa}\right) \ket{0}_j
    \label{eq:exp_approx}
\end{align}
where, $t_{j,\kappa} = \tfrac{\kappa d_j}{1-\kappa d_j}, r_{j,\kappa} = |t_{j,\kappa}|, \theta_{j,\kappa} = \arg(t_{j,\kappa}), \xi_{j,\kappa} = \mathrm{arctanh}(r_{j,\kappa})\exp (i\theta_{j,\kappa}), s_{j,\kappa} = \mathrm{sech}(|\xi_{j,\kappa}|), C_{j,\kappa} = \tfrac{1}{\sqrt{|s_{j,\kappa}|(1-\kappa d_{j})}}\exp \left(\tfrac{(\kappa b_j)^2}{4} +\tfrac{(\kappa b_j) ^2}{2}(1+\xi_{j,\kappa}^*)\xi_{j,\kappa} +\tfrac{|\kappa  b_j(1+\xi_{j,\kappa})|^2}{4}\right)$ and $\alpha_{j,\kappa} = \tfrac{\kappa b_j}{\sqrt{2}}(1+\xi_{j,\kappa})$.
The full derivation can be found in Appendix \ref{app:appendix exp_deriv}. Furthermore, by using the singular value decomposition of $\mathbf{J} = \mathbf{U_2}\mathbf{L} \mathbf{U_1}$, we can obtain the decomposition of the Doktorov operator in terms of Gaussian operators \cite{huh2015,Huh2016VBS}, $\hat U_{\mathrm{Dok}} = \hat{R}(\mathbf{U_2}) \hat S(\ln(\mathbf{L})) \hat{R}(\mathbf{U_1})\hat{D}(\boldsymbol{\beta})$ with $\boldsymbol{\beta} = \tfrac{1}{\sqrt{2}}\mathbf{J}^{-1}\boldsymbol{\delta}$ (where $\ln(\mathbf{L}) = \mathrm{diag}(\ln(\mathbf{l}))$ and $\mathbf{l}$ is the list of $\mathbf{J}$'s singular values).
The complete Gaussian expression of $f_{\mathbf{m^{'}}}$ is as follows:
\begin{widetext}
\begin{align}
    f_{\mathbf{m^{'}}}(\kappa) &= \exp(2\kappa \mu^{(0)})\left|\bra{\mathbf{m^{'}}}\hat U_{\mathrm{Dok}}\hat{R}(\mathbf{U}^{\mathrm{t}})\bigotimes^M_{j=1} C_{j,\kappa}\hat{S}(\xi_{j,\kappa})\hat{D}(\alpha_{j,\kappa}) \ket{0}_j\right|^2 \nonumber \\
    &= \exp(2\kappa \mu^{(0)})\prod^M_{j=1} |C_{j,\kappa}|^2 \left|\bra{\mathbf{m^{'}}}\hat{R}(\mathbf{U_2}) \hat S(\ln(\mathbf{L})) \hat{R}(\mathbf{U_1})\hat{D}(\boldsymbol{\beta})\hat{R}(\mathbf{U}^{\mathrm{t}})\hat{S}(\boldsymbol{\Xi}_\kappa)\hat{D}(\boldsymbol{\alpha}_\kappa) \ket{\mathbf{0}}\right|^2 \nonumber \\
    &= \exp(2\kappa \mu^{(0)})\prod^M_{j=1} |C_{j,\kappa}|^2 \left|\bra{\mathbf{m^{'}}}\hat{R}(\mathbf{V}_\kappa)\hat S(\boldsymbol{\Sigma}_\kappa) \hat{R}(\mathbf{W}_\kappa)^\dagger \hat{D}(\boldsymbol{\gamma}_\kappa)\ket{\mathbf{0}}\right|^2 ,
\end{align}
\end{widetext}
where $\boldsymbol{\alpha}_\kappa = (\alpha_{1,\kappa},\ldots,\alpha_{M,\kappa})^\mathrm{t}$ and $\boldsymbol{\Xi}_\kappa = \mathrm{diag}\left(\xi_{1,\kappa},\ldots,\xi_{M,\kappa} \right)$.
In the third equality, we use the Bloch-Messiah decomposition of the resulting Bogoliubov matrices of the set of Gaussian operators; the derivation and definition of the parameters can be found in Appendix \ref{app:appendix BM decomposition}. Exploiting this decomposition,  $f_{\mathbf{m^{'}}}(\kappa)$ can be prepared experimentally, by preparing  the  $M$-single-mode squeezed coherent state $\hat S(\boldsymbol{\Sigma}_\kappa)\ket{\mathbf{W}_{\kappa }^\mathrm{t}\boldsymbol{\gamma}_\kappa}$ as the input state to the linear optical network characterized by the unitary $\mathbf{V}_\kappa$ and measuring the output photon number distribution. Assuming that $\ket{\mathbf{m^{'}}} = \ket{m_1^{'},\ldots,m_M^{'}}$, the probability of obtaining $m^{'}_i$ photons for the $i^{th}$ output is interpreted as  $\exp(-2\kappa\mu^{(0)})\left(\prod^M_{j=1} |C_{j,\kappa}|^2\right)^{-1}f_{\mathbf{m^{'}}}(\kappa)$.
Finally, the spectrum for the $x$-direction is given by 
\begin{widetext}
\begin{align}
 P_{\tau}(\omega) \simeq \frac{1}{2\mathcal{N}\tau^{2}} \sum_{\mathbf{m^{'}}=\mathbf{0}}^{\boldsymbol{\infty}} \left(f_{\mathbf{m^{'}}}(\mathrm{i}\tau)+\frac{1}{2}f_{\mathbf{m^{'}}}(\tau)+\frac{1}{2}f_{\mathbf{m^{'}}}(-\tau) -2f_{\mathbf{m^{'}}}(0)\right)\delta(\omega-\mathbf{m^{'}}\cdot\boldsymbol{\omega}') , 
\label{eq:vibronic_profile_lin_comb}
\end{align}
\end{widetext}
where the relation is visualized in Fig.~\ref{fig:GBS_vibronic_profile}.

\section{\label{sec:Section num examples} Numerical examples}
As proof of principle, we present  the vibronic spectra of naphthalene and phenanthrene for the first-order HT expansion, and benzene for the first-order and second-order HT expansions. We note that the vibronic spectra of molecules are not the full spectra of the molecules. We considered only a limited number of vibrational modes belonging to a certain symmetry block: a portion of the spectrum that contributes to the whole spectrum. Usually, in the classical simulation, the vibronic profile belonging to each vibrational symmetry block is collected separately, and the entire vibronic profile is obtained by convoluting all vibronic profiles of symmetry blocks. Therefore, if we can obtain the non-Condon profile of a symmetry block with a linear optical device, we can construct the entire vibronic spectrum by convolution. To clearly demonstrate the non-Condon effects, we generated the Franck-Condon profiles to compare the profiles with the non-Condon cases by simply ignoring the TDM operator (i.e., by assuming $\hat{\mu}=1$) and considering only the Doktorov operators.     

\subsection{First-order Herzberg-Teller expansion}

To evaluate our method for the linear HT case, we simulated the ($^{1}\mathrm{B_{2u}}$--$^{1}\mathrm{A_{1g}}$) vibronic transition of naphthalene and the ($^{1}\mathrm{A_{1}}$--$^{1}\mathrm{A_{1}}$) vibronic transition of phenanthrene with two vibrational modes using Strawberry Fields \cite{Killoran2019,hamza_github}, a Python-based quantum optics program package. 
The molecular parameters were extracted from Ref.~\cite{small1971} and can be found in Table \ref{tab:Table 1}.  

\begin{table*}
\caption{\label{tab:Table 1} Molecular parameters for naphthalene and phenanthrene.  $(\mu^{(0)},\mu_1^{(1)},\mu_2^{(1)})$ are the coefficients found in the expansion of the TDM operator, with $\mu^{(0)}$ in D and $\mu_1^{(1)},\mu_2^{(1)}$ in $\mathrm{D}/({\mathrm{u}^{\tfrac{1}{2}}a_0})$. The required parameters were extracted from Ref.~\cite{small1971}, and the TDM was assumed to be polarized only in the $x$-direction. Although the unit of TDM in Ref.~\cite{small1971} was an arbitrary unit, it was assumed to be in Debye without causing problems, as our spectral profiles were normalized. $\boldsymbol{\omega}$ and $\boldsymbol{\omega^{'}}$ are the harmonic frequencies of the initial and excited states, respectively, given in $\mathrm{cm}^{-1}$. $\mathbf{U_\mathrm{D}}$ is the Duschinsky unitary matrix, and $\mathbf{d}$ is the displacement vector given in $\mathrm{u}^{\tfrac{1}{2}}a_0$.}
\begin{ruledtabular}
\begin{tabular}{cccccc}
 Molecule & $(\mu^{(0)},\mu_1^{(1)},\mu_2^{(1)})$ & $\boldsymbol{\omega}$ ($\mathrm{cm}^{-1}$) & $\boldsymbol{\omega^{'}}$  ($\mathrm{cm}^{-1}$) & $\mathbf{U_\mathrm{D}}$ & $\mathbf{d}$ ($\mathrm{u}^{\tfrac{1}{2}}a_0$)\\ \hline
 Naphthalene & $(1.00,1.00,-1.00)$ & $(509.00,938.00)$ & $(438.00,912.00)$ & $\begin{pmatrix}
     0.9800 & -0.2000 \\
     0.2000 & 0.9800
     \end{pmatrix}$ & $(0.0000,0.0000)$ \\ 
     Phenanthrene & $(1.00,1.50,-0.50)$ & $(700.00,800.00)$ & $(679.00,796.00)$ & $\begin{pmatrix}
     0.9055 & -0.4240 \\
     0.4240 &0.9055
     \end{pmatrix}$ & $(0.1650,0.0780)$ \\
\end{tabular}
\end{ruledtabular}
\end{table*}
The transition moment operator for the $x$-direction is given explicitly in this case by, 
\begin{align}
    \hat \mu(\mathbf{\hat{q}}) = \mu^{(0)} + \lambda_1 \hat{q}_1 + \lambda_2 \hat{q}_2 ,
\end{align}
and the auxiliary function $f_{\mathbf{m^{'}}}$ becomes
\begin{align}
    f_{\mathbf{m^{'}}}(\kappa) = \exp(2\kappa \mu^{(0)})\prod^M_{j=1} |C_{j,\kappa}|^2 \left|\bra{\mathbf{m^{'}}}\hat{U}_{\mathrm{Dok}}\hat{D}(\boldsymbol{\alpha}_\kappa) \ket{\mathbf{0}}\right|^2 ,
\end{align}
with $C_{j,\kappa} = \exp\left(\tfrac{|\kappa \lambda_j|^2+(\kappa \lambda_j) ^2}{4}\right) \; \mathrm{and} \;  \alpha_{j,\kappa} = \tfrac{\kappa\lambda_j}{\sqrt{2}}$. 

Because we considered only two vibrational modes of molecules that were relevant to demonstrate our method, this resulted in considering only two modes of the optical device. Moreover, we limited the number of photons per mode up to three ($\mathrm{for} \, \tau=10^{-2}, \,  \sum_{\omega, \, \mathrm{cutoff}=3} P_{\tau}(\omega) \approx 0.9999$ for both molecules), which can be handled by photon-number-resolving detectors.  

\begin{figure}[h!]
  \begin{center}
    \subfloat[]{
      \includegraphics[width=0.45\textwidth,height=0.45\textheight,keepaspectratio]{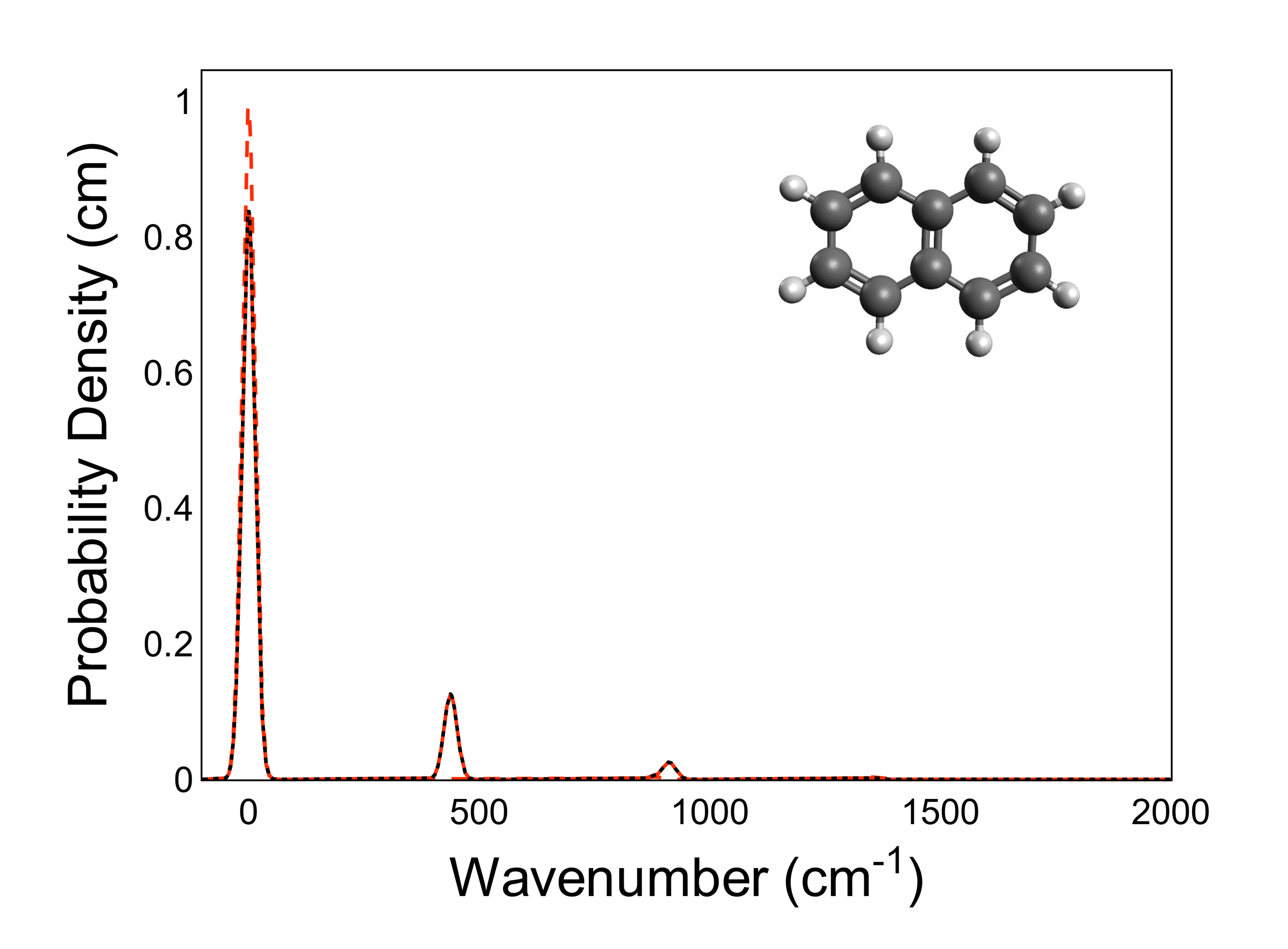}
      \label{sub:Napht_Phen_subfig1}
                         }
                         
    \subfloat[]{
      \includegraphics[width=0.45\textwidth,height=0.45\textheight,keepaspectratio]{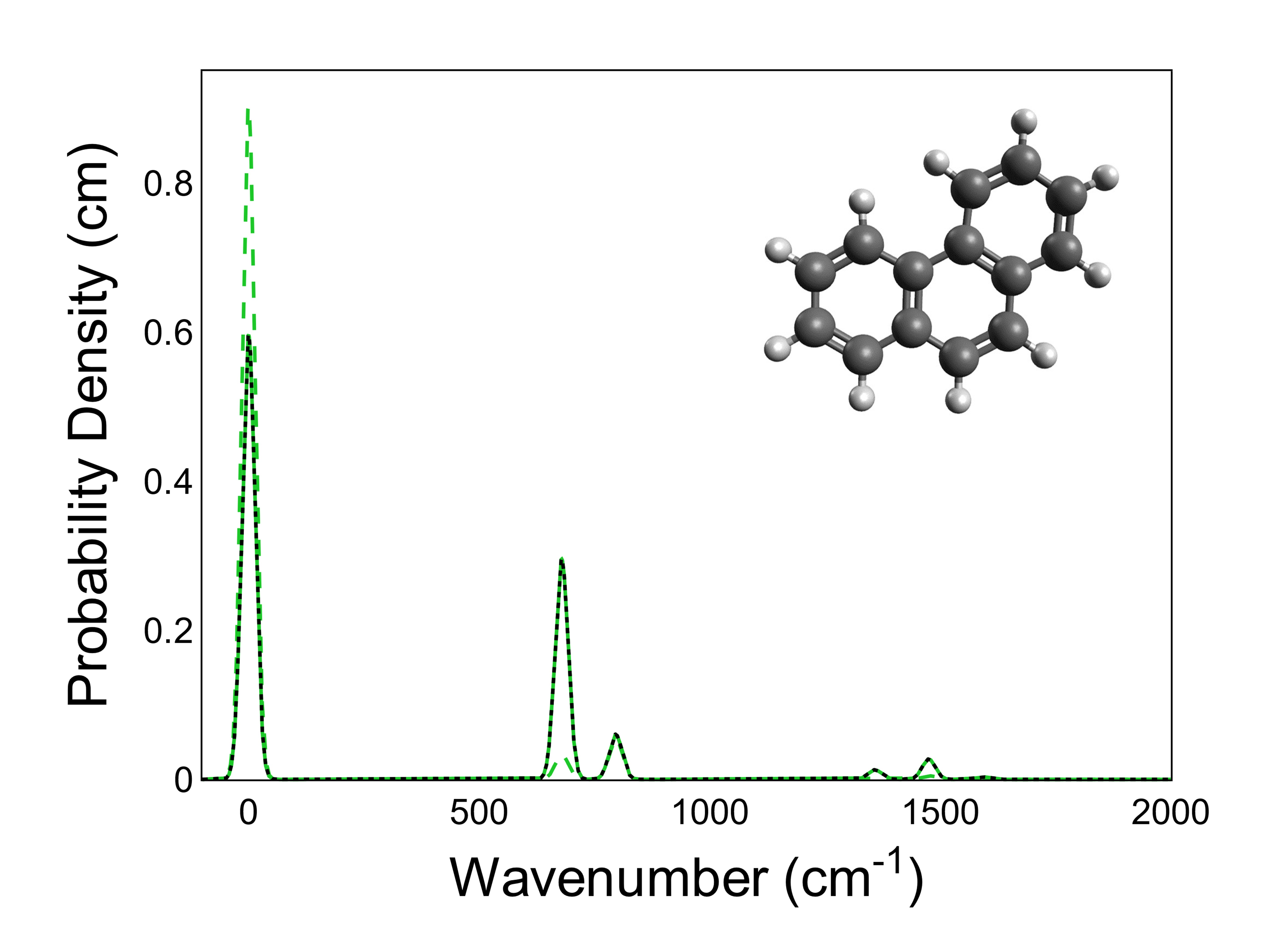}
      \label{sub:Napht_Phen_subfig2}
                         }
    \caption{
    (a) Exact (black solid line), approximate (red dotted line), and Franck-Condon (red dashed line) vibronic spectrum ($\tau = 10^{-2}$) of the vibronic transition of naphthalene  ($^{1}\mathrm{B_{2u}}$--$^{1}\mathrm{A_{1g}}$). 
    (b) Exact (black solid line), approximate (green dotted line), and Franck-Condon (green dashed line) vibronic spectrum ($\tau = 10^{-2}$) of the vibronic transition of phenanthrene ($^{1}\mathrm{A_{1}}$--$^{1}\mathrm{A_{1}}$). Each peak of the spectra was broadened by a Gaussian of width $100 \; \mathrm{cm}^{-1}$.
    }
    \label{fig:Napht_Phen_vibronic_profile}
  \end{center}
\end{figure}

We can calculate the exact spectra (given by Eq.~\eqref{eq:vibronicprofile}) for the two-mode cases by directly evaluate  $\hat{U}_{\mathrm{Dok}}\hat{\mu}(\hat{\mathbf{q}})\ket{\mathbf{0}}$ analytically. The spectra are represented by black solid lines in Fig.~\ref{fig:Napht_Phen_vibronic_profile} for comparison with the approximated spectra. 

As indicated in Table~\ref{tab:Table 1}, the vibronic transition of naphthalene has no displacement. In the Condon regime, therefore, it cannot demonstrate vibronic spectral progression (see red dashed line in  Fig.~\ref{fig:Napht_Phen_vibronic_profile}(a)). In contrast, when we invoke the non-Condon operator, the spectral progression can be obtained as illustrated in Fig.~\ref{fig:Napht_Phen_vibronic_profile}(a). Because $\hat{\mu}(\mathbf{\hat{q}})\ket{00}=\ket{00}+\tfrac{\lambda_1}{\sqrt{2}}\ket{10}+\tfrac{\lambda_2}{\sqrt{2}}\ket{01}$, where $\tfrac{\lambda_1}{\sqrt{2}} = \sqrt{\tfrac{\hbar}{2\omega^{'}_1}}\mu^{(1)}_1 \approx 0.3439$ D and $\tfrac{\lambda_2}{\sqrt{2}} = \sqrt{\tfrac{\hbar}{2\omega^{'}_2}}\mu^{(1)}_2 \approx -0.2533$ D, the linear HT operator effectively provides a displacement operation. As a result, the spectrum has three major peaks at $\omega_{\ket{00}} =  0.00 \, \mathrm{cm}^{-1}$, $\omega_{\ket{10}}=\omega^{'}_1 =438.00\, \mathrm{cm}^{-1}$ and $\omega_{\ket{01}} = \omega^{'}_2 = 912.00 \, \mathrm{cm}^{-1}$. Due to the relatively small Duschinsky rotation, the Doktorov operator only slightly enhances the probability $P(\omega_{\ket{10}})$ and generates small peaks at a high wavenumber domain, causing a peak broadening effect. 

For phenanthrene, we can observe both the linear non-Condon effect and Duschinsky mode mixing effect by comparing the green dashed line and green dotted line in Fig.~\ref{fig:Napht_Phen_vibronic_profile}(b) for the non-Condon and Franck-Condon simulations, respectively. Here, we have $\tfrac{\lambda_1}{\sqrt{2}} = \sqrt{\tfrac{\hbar}{2\omega^{'}_1}}\mu^{(1)}_1 \approx 0.4399$ D and $\tfrac{\lambda_2}{\sqrt{2}} = \sqrt{\tfrac{\hbar}{2\omega^{'}_2}}\mu^{(1)}_2 \approx -0.1372$ D. As the displacement is non-zero, near the peaks caused by the linear HT operator at $ \omega_{\ket{00}} = 0.00, \mathrm{cm}^{-1}$, $\omega_{\ket{10}} = \omega^{'}_1 = 679.00, \mathrm{cm}^{-1}$ and $\omega_{\ket{01}}=\omega^{'}_2 = 796.00, \mathrm{cm}^{-1}$, we have non-negligible neighboring peaks at higher wavenumbers due to the Duschsinky mode mixing effect (See Fig.~\ref{fig:Napht_Phen_vibronic_profile}(b)). 

In Fig. \ref{fig:Napht_Phen_vibronic_profile}, we compare the approximated spectra in dotted lines calculated by the method developed in this paper (Eq.~\eqref{eq:vibronic_profile_lin_comb}) with the exact calculations in solid lines. As indicated in the figure, we can observe an almost perfect match between the spectra of two molecules. This demonstrates the effectiveness of our protocol for simulating the non-unitary operation for the linear HT case. We note that the error (see Fig. \ref{fig:Error_vibronic_profile}) is expected to evolve quadratically according to Eq.~\eqref{eq:linear_comb_aux_func}. Practically, using a value of $\tau = 10^{-1}$ produces a satisfactory result. However, we note here that the smallest $\tau$ will be limited by the quantum device precision: we cannot always increase accuracy simply by lowering $\tau$ unless the quantum device allows it.

Although we obtained matching profiles when limiting the number of modes to two for naphthalene and phenanthrene, it is important to verify the accuracy of our method for a more general case. To this end, we simulated a portion of the vibronic spectrum of benzene corresponding to the $\mathrm{e_{2g}}$ symmetry block (later, in the following section, we examine a smaller symmetry block ($\mathrm{e_{1g}}$) of benzene for the second-order HT expansion). Here, eight vibrational modes were considered, where the number of photons per mode was limited to four ($\mathrm{for} \, \tau=10^{-2}, \,  \sum_{\omega, \, \mathrm{cutoff}=4} P_{\tau}(\omega) \approx 0.9999$), and we considered two components of the TDM operator ($x$ and $y$ directions). The molecular parameters extracted from Ref.~\cite{berger:1998} are provided in Appendix \ref{app:appedix Benzene_e2g_para}. Exact classical simulations were performed using the hotFCHT program package~\cite{berger:1998,jankowiak:2007,Huh2011a,Huh2012} to compare the curve with the approximated curve.  For the two polarization directions, we have  
\begin{align}
    \hat{\mu}_{r}(\mathbf{\hat Q})\ket{\mathbf{0}} &= \sum_{j=1}^{8} \tfrac{\lambda_{r,j}}{\sqrt{2}}\ket{0,\ldots,1_{j},\ldots,0}\nonumber \\
    &= \sum_{j=1}^{8} \sqrt{\tfrac{\hbar}{2\omega_j}}\mu_{r,j}^{(1)}\ket{0,\ldots,1_{j},\ldots,0}
\end{align} 
with $\ket{0,\ldots,1_{j},\ldots,0}$ representing a state with a 1 in the $j^{\mathrm{th}}$ element, where $r \in \{x,y\}$ and,
{\footnotesize
\begin{align}
    &\boldsymbol{\lambda}_{x} = (0.0306,0.0000,0.0000,0.0251,0.0194,0.0000,0.0000,0.1304)^{\mathrm{t}},\nonumber \\
    &\boldsymbol{\lambda}_{y} = (0.0000,0.0306,-0.0251,0.0000,0.0000,-0.0194,-0.1304,0.0000)^{\mathrm{t}}.
\end{align}}
The auxiliary function is given as,
\begin{align}
    f_{r,\mathbf{m^{'}}}(\kappa) = \prod^8_{j=1} |C_{r,j,\kappa}|^2 \left|\bra{\mathbf{m^{'}}}\hat{U}_{\mathrm{Dok}}\hat{D}(\boldsymbol{\alpha}_{r,\kappa}) \ket{\mathbf{0}}\right|^2 
\end{align}
with $C_{r,j,\kappa} = \exp\left(\tfrac{|\kappa \lambda_{r,j}|^2+(\kappa \lambda_{r,j}) ^2}{4}\right) \; \mathrm{and} \;  \alpha_{r,j,\kappa} = \tfrac{\kappa\lambda_{r,j}}{\sqrt{2}}$. 
Finally, the approximated vibronic profile with the TDM operator having two non-zero components is given as,
\begin{widetext}
\begin{align}
P_{\tau}(\omega) \simeq \frac{1}{2\mathcal{N}\tau^{2}} \sum_{r=x,y}\sum_{\mathbf{m^{'}}=\mathbf{0}}^{\boldsymbol{\infty}} \left(f_{r,\mathbf{m^{'}}}(\mathrm{i}\tau)+\frac{1}{2}f_{r,\mathbf{m^{'}}}(\tau)+\frac{1}{2}f_{r,\mathbf{m^{'}}}(-\tau) -2f_{r,\mathbf{m^{'}}}(0)\right)\delta(\omega-\mathbf{m^{'}}\cdot\boldsymbol{\omega}') 
\end{align}
\end{widetext}
where $\mathcal{N} = \sum_{r=x,y} \mathcal{N}_r$. Therefore, we must prepare eight different optical devices to obtain the approximated spectra.

\begin{figure}[h!]
    \begin{center}
    \subfloat[]{
      \includegraphics[width=0.45\textwidth,height=0.45\textheight,keepaspectratio]{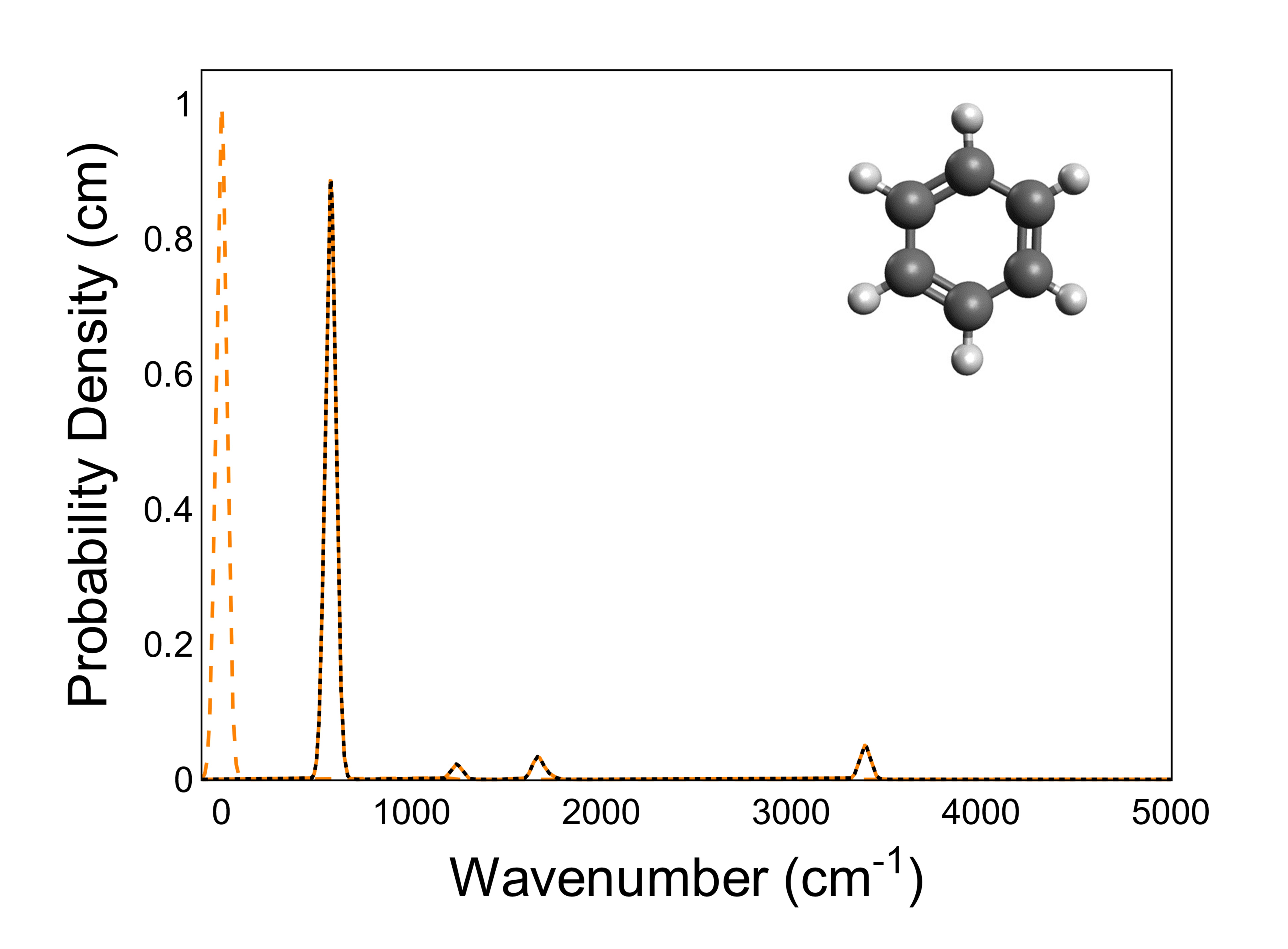}
      \label{sub:Benzene_e2g}
                         }
                         
    \subfloat[]{
      \includegraphics[width=0.45\textwidth,height=0.45\textheight,keepaspectratio]{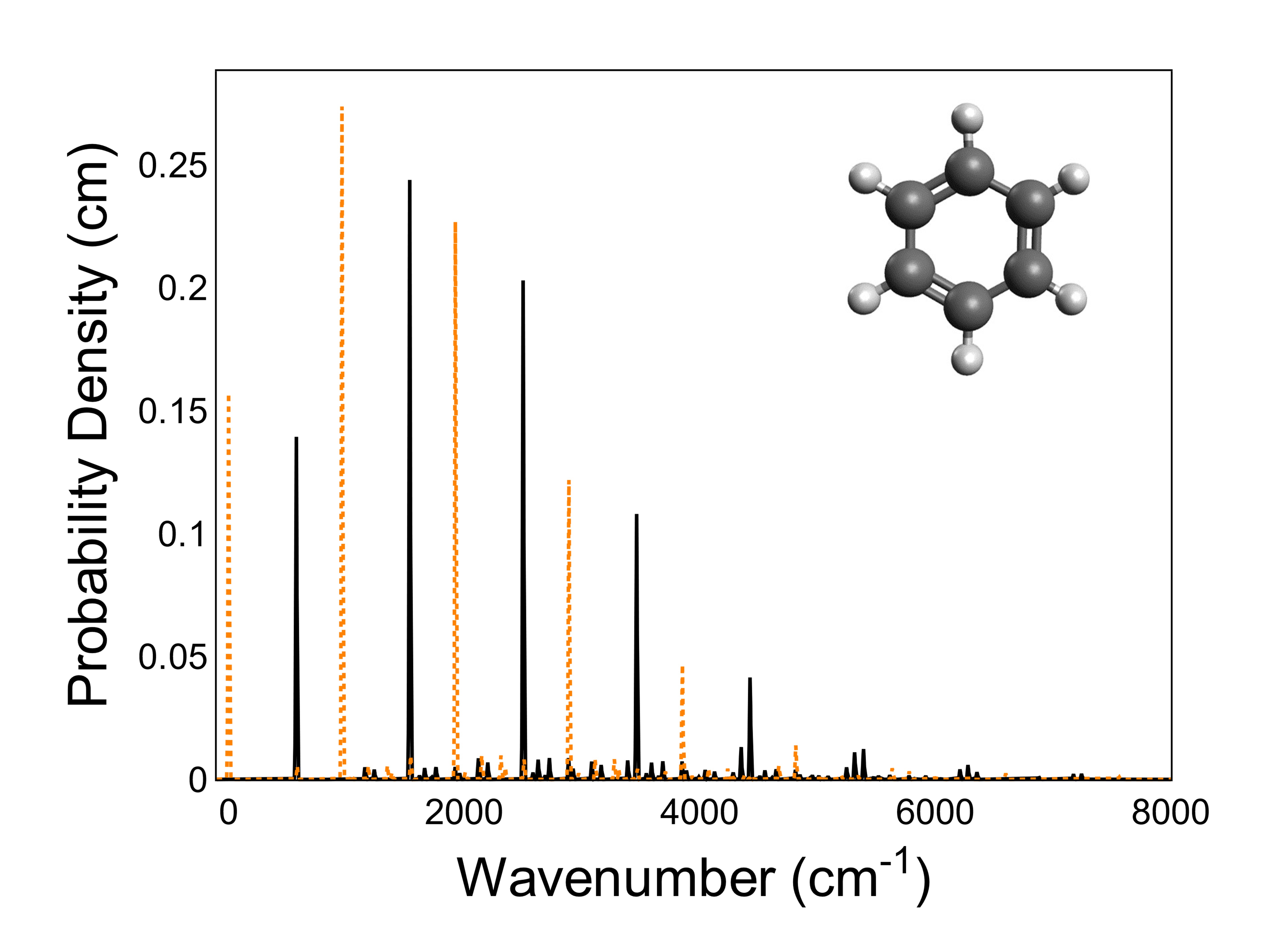}
      \label{sub:Benzene_full}
                         }
    \caption{(a) Exact (black solid line), approximate (orange dotted line), and Franck-Condon (orange dashed line) vibronic spectrum ($\tau = 10^{-2}$) of the vibronic transition of benzene ($^{1}\mathrm{B_{2u}}$--$^{1}\mathrm{A_{1g}}$) restricted to the $\mathrm{e_{2g}}$ symmetry block.
    (b) Full vibronic spectrum of Benzene (black solid line) and full vibronic spectrum where the $\mathrm{e_{2g}}$ symmetry block has been deconvoluted (orange dotted line).
    }
    \label{fig:Benzene_e2g_full}
    \end{center}
\end{figure}

We observed that, for each polarization direction, the dominant coefficient is the one associated with $\omega^{'}_{7} = \omega^{'}_{8} = 575.1367 \; \mathrm{cm}^{-1}$ (eight vibrational modes are doubly degenerated) and as the displacement is small it leads to a dominant peak around $\omega =\omega^{'}_7$. The profile generated by our method and the profile calculated by the classical algorithm plotted in Fig. \ref{fig:Benzene_e2g_full}(a).  agree almost perfectly, which further demonstrates the accuracy of our method. Therein, we also compared the non-Condon spectra with the Franck-Condon curve showing a clear difference. Additionally, we present the full spectrum of benzene (black solid  lines) in Fig. \ref{fig:Benzene_e2g_full}(b), which can be constructed by convolution of partial spectra of symmetry blocks including the spectrum in Fig. \ref{fig:Benzene_e2g_full}(a). 
We additionally deconvoluted the full spectrum with the partial spectrum of e2g symmetry block (black solid lines in Fig.~\ref{fig:Benzene_e2g_full}(a)) and present it in Fig.~\ref{fig:Benzene_e2g_full}(b) as orange dashed lines to show the non-Condon effects clearly.

\subsection{Second-order Herzberg-Teller expansion}
In the previous numerical section, we present an evaluation of our method for linear HT cases. In this subsection, we describe an evaluation of our method for second-order HT expansion with two vibrational modes of benzene belonging to a different symmetry block $\mathrm{e_{1g}}$. 
By collecting information about the $\mathrm{e_{1g}}$ symmetry block of the vibronic transition of benzene ($^{1}\mathrm{B_{2u}}$--$^{1}\mathrm{A_{1g}}$)~\cite{fischerbook,fischer1981}, we created the following model to evaluate the proposed method for the second-order HT expansion. As this transition is Franck-Condon-forbidden ($\mu^{(0)}=0$), it is an excellent candidate to illustrate our method. The molecular parameters are summarized in Table \ref{tab:Table 2}.
\begin{table*}
\caption{\label{tab:Table 2} Molecular parameters for three $\mathrm{e_{1g}}$ vibrational modes of benzene. $\boldsymbol{\omega}$ and $\boldsymbol{\omega^{'}}$ are the harmonic frequencies of the initial and excited states, respectively,  given in $\mathrm{cm}^{-1}$. $\mathbf{U_\mathrm{D}}$ is the Duschinsky unitary matrix. The displacement is zero in this case. The elements in $\mathbf{M}$ are given in $\mathrm{D}/({\mathrm{u}^{\tfrac{1}{2}}a_0})^2$.  }
\begin{ruledtabular}
\begin{tabular}{ccccc}
 Molecule & $\mathbf{M}$ & $\boldsymbol{\omega}$ ($\mathrm{cm}^{-1}$) & $\boldsymbol{\omega^{'}}$  ($\mathrm{cm}^{-1}$) & $\mathbf{U_\mathrm{D}}$\\ \hline
 Benzene & $\begin{pmatrix} 0 & \mu^{(2)}_{1,2} & \mu^{(2)}_{1,2} \\
    \mu^{(2)}_{1,2} & \mu^{(2)}_{2,2} & \mu^{(2)}_{2,2} \\
    \mu^{(2)}_{1,2} & \mu^{(2)}_{2,2} & -\mu^{(2)}_{2,2} 
    \end{pmatrix}$\footnote{$\mu^{(2)}_{1,2} = 0.0463 \mathrm{D}/({\mathrm{u}^{\tfrac{1}{2}}a_0})^2$ and $\mu^{(2)}_{2,2} = 0.0216\mathrm{D}/({\mathrm{u}^{\tfrac{1}{2}}a_0})^2$}  & $(712.6271,869.5370,869.5370)$ & $(482.2731,593.2363,593.2363)$ & $ \begin{pmatrix}
     1 &0 &0 \\
     0&0 & 1 \\
 0&1 & 0\\
 \end{pmatrix}$ \\ 
\end{tabular}
\end{ruledtabular}
\end{table*}
The transition moment operator is given by
\begin{align} 
    \hat \mu (\hat{\mathbf{q}}) =  \hat{\mathbf{q}}^{\mathrm{t}} \boldsymbol{\Lambda} \hat{\mathbf{q}} ,
\end{align}
leading to,
\begin{align}
    f_{\mathbf{m^{'}}}(\kappa) = \prod^M_{j=1} |C_{j,\kappa}|^2 \left|\bra{\mathbf{m^{'}}}\hat{U}_{\mathrm{Dok}}\hat{R}(\mathbf{U}^{\mathrm{t}})\hat{S}(\boldsymbol{\Xi}_\kappa) \ket{\mathbf{0}}\right|^2 ,
\end{align}
with $C_{j,\kappa} = \tfrac{1}{\sqrt{|s_{j,\kappa}|(1-\kappa d_{j})}} \; \mathrm{and} \; \boldsymbol{\Xi}_\kappa = \mathrm{diag}\left(\xi_{1,\kappa},\ldots,\xi_{M,\kappa} \right)$ defined in (\ref{eq:exp_approx}).

\begin{figure}[h!]
    \centering
    \includegraphics[width=0.45\textwidth,height=0.45\textheight,keepaspectratio]{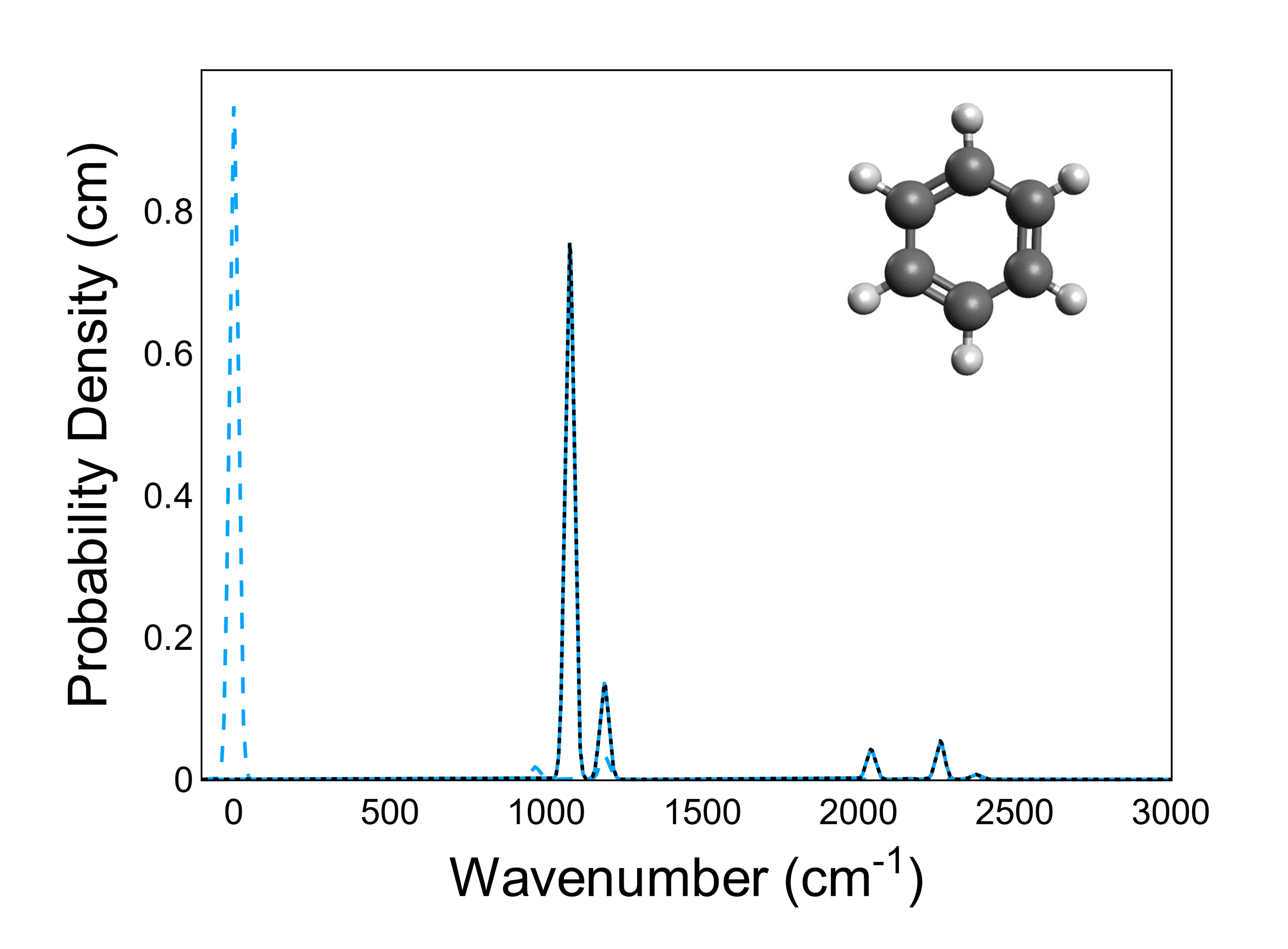}
    \caption{\label{fig:Benz_vibronic_profile} Exact (black solid line), approximate (blue dotted line), and Franck-Condon (blue dashed line) vibronic spectrum ($\tau = 10^{-2}$) of the vibronic transition of benzene ($^{1}\mathrm{B_{2u}}$--$^{1}\mathrm{A_{1g}}$). }
\end{figure}

Here, we used a three-mode model and limited the number of photons per mode to five ($\mathrm{for} \, \tau=10^{-2}, \,  \sum_{\omega, \, \mathrm{cutoff}=5} P_{\tau}(\omega) \approx 0.9993$). This limitation will not hinder the experiment, as the photon-number-resolving detection scheme developed in \cite{wang2019} could resolve up to $n=16$ Fock states per mode. However, we could recover large portions of the profile with a cutoff of three photons per mode ($\mathrm{for} \, \tau=10^{-2}, \,  \sum_{\omega, \, \mathrm{cutoff}=3} P_{\tau}(\omega) \approx 0.9872$), leading to a simplified experiment. The results are plotted in Fig. \ref{fig:Benz_vibronic_profile}.
The Duschinsky rotation is an identity matrix (by a proper permutation), and because there is no displacement, we must find a spectrum close to   $\hat{U}_{\mathrm{Dok}}\hat{\mu}(\hat{\mathbf{q}})\ket{000}\approx\Lambda_{1,2}(\ket{110}+\ket{101})+\Lambda_{2,2}(\ket{011}+\tfrac{1}{\sqrt{2}}\ket{020}+\tfrac{1}{\sqrt{2}}\ket{002})$. As $\omega^{'}_2 =\omega^{'}_3$ (meaning that the permutation does not do anything), we have $\omega_{\ket{110}} = \omega_{\ket{101}}$ and $\omega_{\ket{011}} = \omega_{\ket{020}} = \omega_{\ket{002}}$. We must then obtain a peak at $\omega_{\ket{110}} = \omega^{'}_1 + \omega^{'}_2 \approx 1.0760 \times 10^{3} \, \mathrm{cm}^{-1}$ and another at $\omega_{\ket{011}} = 2\omega^{'}_2 \approx 1.1860 \times 10^{3} \, \mathrm{cm}^{-1}$, with $P(\omega_{\ket{110}}) = \tfrac{\Lambda_{1,2}^2}{\Lambda_{1,2}^2+\Lambda_{2,2}^2} \approx 0.8486$ and $P(\omega_{\ket{011}}) = \tfrac{\Lambda_{2,2}^2}{\Lambda_{1,2}^2+\Lambda_{2,2}^2} \approx 0.1514$. Peaks at higher wavenumbers come from $\hat{U}_{\mathrm{Dok}}$ and induce a small reduction of the precedent probabilities. Here, we also present the Franck-Condon spectum (dashed blue line in Fig.~\ref{fig:Benz_vibronic_profile}) to compare it with the non-Condon spectra. 

\begin{figure}[h!]
    \centering
    \includegraphics[width=0.45\textwidth,height=0.45\textheight,keepaspectratio]{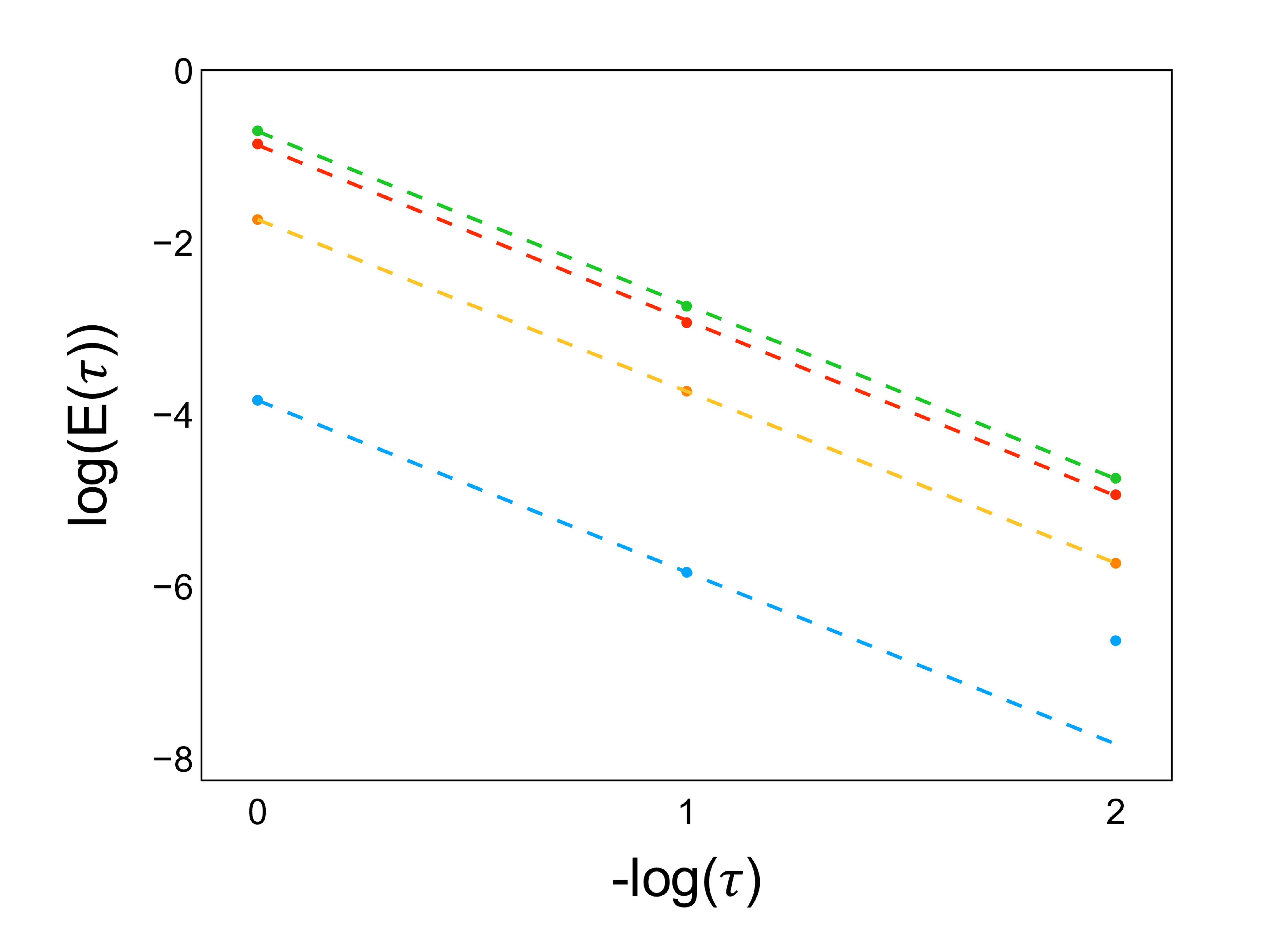}
    \caption{\label{fig:Error_vibronic_profile} Error between exact and approximate profiles for different values of $\tau$ ($\mathrm{E}(\tau) = \vert P_{\mathrm{exact}}-P_{\tau}\vert$).  
    The red dashes represent the error for naphthalene, the green dashes the error for phenanthrene, and the yellow and blue dashes represent the first-order and second-order HT transion of benzene, respectively. 
    The slopes of the lines are close to -2, which is expected from the method developed in this paper. 
    }
\end{figure}

Lastly, we analyzed the errors depending on the size of the parameter $\tau$ (see Fig.~\ref{fig:Error_vibronic_profile}). Ideally, the error behaves quadratically with $\tau$. As indicated in  Fig.~\ref{fig:Error_vibronic_profile}, almost all cases exhibited ideal behavior. 
However, one error of benzene (second-order HT expansion at $\tau = 10^{-2}$) deviated from the expected value. To examine the cause, we consider an effective parameter $\tau_{\mathrm{eff}}$. For benzene, $\hat{\mu} = \hat{\mathbf{q}}^{\mathrm{t}}\boldsymbol{\Lambda}\hat{\mathbf{q}}$ where all the parameters in $\boldsymbol{\Lambda}$ are on the order of $10^{-3}$ (after conversion). We can then rewrite $\hat{\mu} = 10^{-3}\hat{\mu}_{\mathrm{eff}}$, which leads to rescaling the expansion parameter $\tau$ effectively, that is, $\tau_{\mathrm{eff}} = 10^{-3}\tau$. Therefore, taking a small value of $\tau$ in the expansion, we would encounter a numerical rounding-off error earlier than in the linear case.  
\section{Conclusion and outlook}
Without consideration of practical applications, boson sampling was introduced as an easily implementable device whose computational power can surpass that of classical computers. However, the proposal by Huh et al. \cite{huh2015} revealed that boson sampling can also be useful for quantum simulation. Our work generalizes the aforementioned study by simulating more complex molecular processes beyond the Condon regime, thereby increasing the number of practical applications of boson sampling \cite{bromley_2020}. The non-Condon profile was  approximated with linear optics simply by using a linear combination of Gaussian boson samplers that we introduced. We note here that our new development is not just an extension of the theoretical proposal of Huh and coworkers~\cite{huh2015}. The method opened a new way to approximate non-unitary operations in a linear optical network that it can be generalized to arbitrary non-unitary operators and digital quantum simulations. Our technique can potentially enable the extraction of relevant molecular properties that exhibit strong non-Condon effects without resorting to the use of digital quantum computers for the Condon case~\cite{sawaya2019,sawaya2020_dlev,sawaya2020_spec}.

Focusing on small molecules to motivate experimental implementations, we evaluated our method for both linear and quadratic HT expansions through numerical examples. In each case, we successfully simulated the non-Condon profiles with a small error. However, because we studied probability densities, we did not consider errors due to sampling. In fact, we had infinite precision in the estimation of the probabilities. In an experiment in which the profile is reconstructed by sampling from the approximated probability density, there will be errors due to the finite number of samples considered.

Our work can be extended to other types of vibronic transitions such as resonance Raman spectroscopy~\cite{santoro2011}, internal conversion~\cite{peng2007}, and intersystem crossing~\cite{ETINSKI2011}, as all of these are related by the computation of a non-Condon integral to evaluate the probability of a transition. 
Furthermore, we can also generalize the work performed in \cite{sawaya2019} for the quantum circuit model to include non-Condon effects based on the current development or to describe the non-adiabatic molecular quantum dynamics~\cite{Ollitrault2020-1}. Finally, we can reuse the concepts developed in this paper to improve current classical algorithms used for molecular vibronic spectroscopy in the non-Condon case; for example, we can combine the current approach with the time-dependent method~\cite{Huh2011a,baiardi:2013}.

\begin{acknowledgments}
	This work is supported by Basic Science Research Program through the National Research Foundation of Korea (NRF) funded by the Ministry of Education, Science and Technology (NRF-2015R1A6A3A04059773,  NRF-2019M3E4A1080227, NRF-2019M3E4A1079666). JH acknowledges the support by the POSCO Science Fellowship of POSCO TJ Park Foundation. 
\end{acknowledgments}

\appendix
\begin{widetext}

\section{\label{app:appendix Norm_const} Computation of the normalization constant}
We start from the normalization condition,
\begin{align}
    1 =\sum_{\omega=0}^{\infty} P(\omega)=\frac{1}{\mathcal{N}}\sum_ {r=x,y,z}\sum_{\omega=0}^{\infty}\sum_{\mathbf{m^{'}}=\mathbf{0}}^{\boldsymbol{\infty}}\vert\langle\mathbf{m^{'}}\vert\hat{U}_{\mathrm{Dok}}\hat{\mu}_{r}\vert\mathbf{0}\rangle\vert^{2}\delta(\omega-\mathbf{m^{'}}\cdot\boldsymbol{\omega}') .
\end{align}
But,
\begin{align}
    \sum_{\omega=0}^{\infty}\sum_{\mathbf{m^{'}}=\mathbf{0}}^{\boldsymbol{\infty}}\vert\langle\mathbf{m^{'}}\vert\hat{U}_{\mathrm{Dok}}\hat{\mu}_{r}\vert\mathbf{0}\rangle\vert^{2}\delta(\omega-\mathbf{m^{'}}\cdot\boldsymbol{\omega}') &= \sum_{\omega=0}^{\infty}\sum_{\mathbf{m^{'}}=\mathbf{0}}^{\boldsymbol{\infty}}\bra{\mathbf{0}}\hat{\mu}_{r}^{\dagger}\hat U_{\mathrm{Dok}}^{\dagger}\ket{\mathbf{m^{'}}}\bra{\mathbf{m^{'}}}\hat U_{\mathrm{Dok}} \hat{\mu}_{r} \ket{\mathbf{0}}\delta(\omega-\mathbf{m^{'}}\cdot\boldsymbol{\omega}') \nonumber \\
    &= \bra{\mathbf{0}}\hat{\mu}_{r}^{\dagger}\hat{U}_{\mathrm{Dok}}^{\dagger}\left(\sum_{\omega=0}^{\infty}\sum_{\mathbf{m^{'}}=\mathbf{0}}^{\boldsymbol{\infty}}\ket{\mathbf{m^{'}}}\bra{\mathbf{m^{'}}}\delta(\omega-\mathbf{m^{'}}\cdot\boldsymbol{\omega}')\right)\hat{U}_{\mathrm{Dok}}\hat{\mu}_{r}\ket{\mathbf{0}} .
\end{align}
And,
\begin{align}
    \sum_{\omega=0}^{\infty}\sum_{\mathbf{m^{'}}=\mathbf{0}}^{\boldsymbol{\infty}}\ket{\mathbf{m^{'}}}\bra{\mathbf{m^{'}}}\delta(\omega-\mathbf{m^{'}}\cdot\boldsymbol{\omega}')&= \sum_{\mathbf{m^{'}}=\mathbf{0}}^{\boldsymbol{\infty}}\ket{\mathbf{m^{'}}}\bra{\mathbf{m^{'}}}\left(\sum_{\omega=0}^{\infty}\delta(\omega-\mathbf{m^{'}}\cdot\boldsymbol{\omega}')\right) \nonumber \\
    &= \sum_{\mathbf{m^{'}}=\mathbf{0}}^{\boldsymbol{\infty}}\ket{\mathbf{m^{'}}}\bra{\mathbf{m^{'}}} \nonumber \\
    &= \mathcal{I}.
\end{align} 
Moreover $\hat{\mu}_{r} = \hat{\mu}_{r}^{\dagger}$ which leads to,
\begin{align}
    \mathcal{N} = \sum_{r=x,y,z}\mathcal{N}_{r} = \sum_{r=x,y,z}\bra{\mathbf{0}}\hat{\mu}_{r}\hat{U}_{\mathrm{Dok}}^{\dagger}\hat{U}_{\mathrm{Dok}}\hat{\mu}_{r}\ket{\mathbf{0}} = \sum_{r=x,y,z}\bra{\mathbf{0}}\hat{\mu}_{r}^2\ket{\mathbf{0}}
\end{align}
as $\hat{U}_{\mathrm{Dok}}$ is a unitary operator. 
We derive the normalization factor for the second-order expansion, the first-order one being found by taking $\boldsymbol{\Lambda}_{r} = \mathbf{0}$. We have, $\hat{\mu}_{r}(\mathbf{\hat q})=\mu_{r}^{(0)}
+\sum_{j=1}^{M}\lambda_{r,j}\hat{q}_{r,j}+\sum_{j,k=1}^{M}[\boldsymbol{\Lambda}_{r}]_{j,k}\hat{q}_{r,j}\hat{q}_{r,k}$. When considering the overlap $\bra{\mathbf{0}}\hat{\mu}_{r}^2\ket{\mathbf{0}}$, the terms of the type $\bra{\boldsymbol{0}}\hat q_{r,j} \hat q_{r,k}\ket{\boldsymbol{0}}$ for $j \neq k$, or $\bra{\boldsymbol{0}}\hat q_{r,i} \hat q_{r,j} \hat q_{r,k} \hat q_{r,l}\ket{\boldsymbol{0}}$ for $\{k,l\}$ not a permutation of $\{i,j\}$ vanish as we get the overlap between two orthogonal basis vectors. Finally, we get,
\begin{align}
    \mathcal{N}_{2} &= \sum_{r=x,y,z}\mathcal{N}_{r,2} \nonumber \\
    &= \sum_{r=x,y,z} \left((\mu_{r}^{(0)})^2 + \tfrac{1}{2}\sum_{j=1}^M\lambda_{r,j}^2 +\mu_{r}^{(0)}\sum_{j=1}^{M}[\boldsymbol{\Lambda}_{r}]_{j,j}+ \tfrac{1}{4}\sum_{\substack{j,k=1\\ j \neq  k}}^{M}[\boldsymbol{\Lambda}_{r}]_{j,j}[\boldsymbol{\Lambda}_{r}]_{k,k}+ \tfrac{1}{2}\sum_{\substack{j,k=1\\ j \neq  k}}^{M} [\boldsymbol{\Lambda}_{r}]_{j,k}^2 + \tfrac{3}{4}\sum_{j=1}^{M} [\boldsymbol{\Lambda}_{r}]_{j,j}^2\right) . 
\end{align}

\section{\label{app:appendix exp_deriv}Derivation of $\exp(\kappa\hat{\mu})\vert\mathbf{0}\rangle$}

If $\kappa$ was purely imaginary, then $\exp(\kappa \hat\mu)$ would be a unitary (as $\hat \mu$ is Hermitian) and it would be straightforward to implement the linear optical setup. However, it is necessary to generalize the application of the operator for $\kappa$ can be a real number. Therefore, we have, 
\begin{align}
\exp(\kappa\hat{\mu})\vert\mathbf{0}\rangle
&=\hat{R}(\mathbf{U}^\mathrm{t})\exp(\kappa(\mu^{(0)}
+\mathbf{b}^{\mathrm{t}}\hat{\mathbf{q}}
+\hat{\mathbf{q}}^{\mathrm{t}}\mathbf{D}\hat{\mathbf{q}}))\vert\mathbf{0}\rangle
\nonumber \\
&=\exp(\kappa\mu^{(0)})\hat{R}(\mathbf{U}^\mathrm{t})\bigotimes_{j=1}^{M}\exp(\kappa b_{j}\hat{q}_{j}+\kappa d_{j}\hat{q}_{j}^{2})\vert 0\rangle_{j} . \nonumber \\
\end{align}
Now, we need to find the expression of $\exp(\kappa b_{j}\hat{q}_{j}+\kappa d_{j}\hat{q}_{j}^{2})\vert 0\rangle_{j} = \exp(\kappa b_{j}\hat{q}_{j})\exp(\kappa d_{j}\hat{q}_{j}^{2})\vert 0\rangle_{j} $ (as $[\hat q_j,\hat q_j^2] = 0$) in terms of Gaussian operators.
On the one hand~\cite{Fan2003},
\begin{align}
    \exp(\kappa d_{j}\hat{q}_{j}^{2})\ket{0}_j
    &=\int_{-\infty}^{\infty}\mathrm{d}q_{j} \exp(\kappa d_{j}q_{j}^{2})\vert q_{j}\rangle\langle q_{j}\vert \ket{0}_j\nonumber \\
    &=\int_{-\infty}^{\infty}\frac{\mathrm{d}q_{j}}{\sqrt{\pi}} \exp(\kappa d_{j}q_{j}^{2}):\exp(-(q_{j}-\hat{q}_{j})^{2}):\ket{0}_j    \nonumber \\
    &=\int_{-\infty}^{\infty}\frac{\mathrm{d}q_{j}}{\sqrt{\pi}} :\exp\left(-(1-\kappa d_{j})q_{j}^{2}+2\hat{q}_{j}q_{j}-\hat{q}_{j}^{2}\right):\ket{0}_j\nonumber \\
    &=\frac{1}{\sqrt{1-\kappa d_{j}}}:\exp\left(\frac{\kappa d_j}{1-\kappa d_j}\hat{q}_{j}^{2}\right):\ket{0}_j \nonumber \\
    &= \frac{1}{\sqrt{1-\kappa d_{j}}}\exp\left(\frac{1}{2}\frac{\kappa d_j}{1-\kappa d_j}\hat{a}_{j}^{\dagger2}\right)\ket{0}_j ,
\end{align}
where : : denotes normal ordering. On the other hand, the squeezing operator can be arranged in a normal ordered form,
\begin{align}
    \hat{S}_{j}(\xi_{j})\ket{0}_j&=\exp(\tfrac{1}{2}(\xi_{j}\hat{a}_{j}^{\dagger 2}-\xi_{j}^{*}\hat{a}_{j}^{2}))\ket{0}_j\nonumber \\
    &=\vert s_{j}\vert^{\tfrac{1}{2}}\exp(\tfrac{1}{2}t_{j}\hat{a}_{j}^{\dagger 2})
    :\exp((s_{j}-1)\hat{a}_{j}^{\dagger}\hat{a}_{j}):
    \exp(-\tfrac{1}{2}t_{j}\hat{a}_{j}^{2})\ket{0}_j \nonumber \\
    & = \vert s_{j}\vert^{\tfrac{1}{2}}\exp(\tfrac{1}{2}t_{j}\hat{a}_{j}^{\dagger 2})\ket{0}_j ,
\end{align}
where $\xi_{j}=r_{j}\exp(\mathrm{i}\theta_{j})$, $t_{j}=\tanh(r_{j})\exp(\mathrm{i}\theta_{j})$, $s_{j}=\mathrm{sech}(r_{j})$. Then by identification, 

\begin{align}
    \exp(\kappa d_{j}\hat{q}_{j}^{2})\ket{0}_j = \frac{1}{\sqrt{|s_{j,\kappa}|(1-\kappa d_{j})}}\hat{S}_{j}(\xi_{j,\kappa})\ket{0}_j .
\end{align}
with $t_{j,\kappa} = \tfrac{\kappa d_j}{1-\kappa d_j}$. $r_{j,\kappa}, \theta_{j,\kappa}$ and $\xi_{j,\kappa}$ stem from $t_{j,\kappa}$. Then, 
\begin{align}
    \exp(\kappa b_{j}\hat{q}_{j}+\kappa d_{j}\hat{q}_{j}^{2})\ket{0}_j &= \exp(\kappa b_{j}\hat{q}_{j})\exp(\kappa d_{j}\hat{q}_{j}^{2})\ket{0}_j \nonumber \\
    &= \frac{1}{\sqrt{|s_{j,\kappa}|(1-\kappa d_{j})}}\exp(\kappa b_{j}\hat{q}_{j})\hat{S}_{j}(\xi_{j,\kappa})\ket{0}_j
\end{align}
By using the formulas $\exp(A)\exp(B) = \exp(B)\exp(A)\exp([A,B])$ and $\exp(A+B) = \exp(A)\exp(B)\exp(-\tfrac{1}{2}[A,B])$, we derive the final expression of $\exp(\kappa\hat\mu)\ket{\mathbf{0}}$,
\begin{align}
&\exp(\kappa b_{j}\hat{q}_{j}+\kappa d_{j}\hat{q}_{j}^{2})\ket{0}_j \nonumber \\
&=\exp(\kappa b_{j}\hat{q}_{j})\exp(\kappa d_{j}\hat{q}_{j}^{2})\vert 0\rangle_{j} \nonumber \\
&= \frac{1}{\sqrt{|s_{j,\kappa}|(1-\kappa d_{j})}}\exp \left(\frac{(\kappa b_j)^2}{4}\right) \exp\left(\frac{\kappa b_j}{\sqrt{2}}\hat a_j^{\dagger}\right)\exp\left(\frac{\kappa b_j}{\sqrt{2}} \hat a_j\right) \hat S_j(\xi_{j,\kappa}) \ket{0}_j \nonumber \\
&= \frac{1}{\sqrt{|s_{j,\kappa}|(1-\kappa d_{j})}}\exp \left(\frac{(\kappa b_j)^2}{4}\right) \exp\left(\frac{\kappa b_j}{\sqrt{2}}\hat a_j^{\dagger}\right) \hat S_j(\xi_{j,\kappa})\exp\left(\frac{\kappa b_j}{\sqrt{2}} \hat a_j\right)\exp\left(\frac{\kappa b_j \xi_{j,\kappa}}{\sqrt{2}} \hat a_j^{\dagger}\right) \ket{0}_j \nonumber \\
&= \frac{1}{\sqrt{|s_{j,\kappa}|(1-\kappa d_{j})}}\exp \left(\frac{(\kappa b_j)^2}{4}\right)  \hat S_j(\xi_{j,\kappa})\exp\left(\frac{\kappa b_j}{\sqrt{2}} \hat a_j^{\dagger}\right)\exp\left(\frac{\kappa b_j }{\sqrt{2}}(1+\xi_{j,\kappa}^*) \hat a_j\right)\exp\left(\frac{\kappa b_j \xi_{j,\kappa}}{\sqrt{2}} \hat a_j^{\dagger}\right) \ket{0}_j \nonumber \\
&=\frac{1}{\sqrt{|s_{j,\kappa}|(1-\kappa d_{j})}}\exp \left(\frac{(\kappa b_j)^2}{4} + \frac{(\kappa b_j)^2}{2} (1+\xi_{j,\kappa}^*) \xi_{j,\kappa}\right)  \hat S_j(\xi_{j,\kappa})\exp\left(\frac{\kappa b_j}{\sqrt{2}}(1+\xi_{j,\kappa}) \hat a_j^{\dagger}\right)\exp\left(\frac{\kappa b_j }{\sqrt{2}}(1+\xi_{j,\kappa}^*)\hat a_j\right) \ket{0}_j . \nonumber \\
\end{align}
However, $\exp\left(\tfrac{\kappa b_j }{\sqrt{2}}(1+\xi_{j,\kappa}^*)\hat a_j\right) \ket{0}_j = \ket{0}_j$ because $\hat a_j \ket{0}_j = 0$ and,
\begin{align}
    &\exp\left(\frac{\kappa b_j}{\sqrt{2}}(1+\xi_{j,\kappa}) \hat a_j^{\dagger}\right)\ket{0}_j \nonumber \\
    &= \exp\left(\frac{\kappa b_j}{\sqrt{2}}(1+\xi_{j,\kappa}) \hat a_j^{\dagger}\right) \exp\left(\left(\frac{\kappa b_j}{\sqrt{2}}(1+\xi_{j,\kappa})\right)^{*}\hat a_j\right) \ket{0}_j \nonumber \\ &= \exp\left(\frac{|\kappa b_j(1+\xi_{j,\kappa})|^2 }{4}\right)\hat D_j \left(\frac{\kappa b_j}{\sqrt{2}}(1+\xi_{j,\kappa})\right)\ket{0}_j ,
\end{align}
which leads to,
\begin{align}
    &\exp(\kappa b_{j}\hat{q}_{j}+\kappa d_{j}\hat{q}_{j}^{2})\ket{0}_j \nonumber \\
    &= \frac{1}{\sqrt{|s_{j,\kappa}|(1-\kappa d_{j})}}\exp \left(\frac{(\kappa b_j)^2}{4} + \frac{(\kappa b_j) ^2}{2}(1+\xi_{j,\kappa}^*)\xi_{j,\kappa} +\frac{|\kappa b_j(1+\xi_{j,\kappa})|^2}{4}\right) \hat S_j(\xi_{j,\kappa})\hat D_j\left(\frac{\kappa b_j (1+\xi_{j,\kappa})}{\sqrt{2}}\right) \ket{0}_j \nonumber \\
    &= C_{j,\kappa} \hat S_j(\xi_{j,\kappa})\hat D_j\left(\alpha_{j,\kappa} \right) \ket{0}_j ,
\end{align}
where $t_{j,\kappa} = \tfrac{\kappa d_j}{1-\kappa d_j}$, $ r_{j,\kappa} = |t_{j,\kappa}|$, $ \theta_{j,\kappa} = \arg(t_{j,\kappa})$, $\xi_{j,\kappa} = \mathrm{arctanh}(r_{j,\kappa})\exp(i\theta_{j,\kappa})$, $s_{j,\kappa} = \mathrm{sech}(|\xi_{j,\kappa}|)$, $C_{j,\kappa} = \tfrac{1}{\sqrt{|s_{j,\kappa}|(1-\kappa d_{j})}}\exp \left(\tfrac{(\kappa b_j)^2}{4} + \tfrac{(\kappa b_j) ^2}{2}(1+\xi_{j,\kappa}^*)\xi_{j,\kappa} +\tfrac{|\kappa b_j(1+\xi_{j,\kappa})|^2}{4}\right)$ and $\alpha_{j,\kappa} = \tfrac{\kappa b_j}{\sqrt{2}}(1+\xi_{j,\kappa})$.

\section{\label{app:appendix BM decomposition} Bloch-Messiah decomposition}
Initially we have,
\begin{align}
    f_{\mathbf{m^{'}}}(\kappa) = \exp(2\kappa \mu^{(0)})\prod^M_{j=1} |C_{j,\kappa}|^2 \left|\bra{\mathbf{m^{'}}}\hat{R}(\mathbf{U_2}) \hat S(\mathbf{\ln(L)}) \hat{R}(\mathbf{U_1})\hat{D}(\boldsymbol{\beta})\hat{R}(\mathbf{U}^{\mathrm{t}})\hat{S}(\boldsymbol{\Xi}_\kappa)\hat{D}(\boldsymbol{\alpha}_\kappa) \ket{\mathbf{0}}\right|^2 
\end{align}
However, by using the Bloch-Messiah decomposition we can rewrite \cite{Braunstein2005,Cariolaro2016} the Gaussian operator $\hat{R}(\mathbf{U_2}) \hat S(\mathbf{\ln(L)}) \hat{R}(\mathbf{U_1})\hat{D}(\boldsymbol{\beta})\hat{R}(\mathbf{U}^{\mathrm{t}})\hat{S}(\boldsymbol{\Xi}_\kappa)\hat{D}(\boldsymbol{\alpha}_\kappa)$ in a simpler form, which will lead to a simpler experimental realization. 
Let $\hat O_\kappa =\hat{R}(\mathbf{U_2}) \hat S(\mathbf{\ln(L)}) \hat{R}(\mathbf{U_1})\hat{D}(\boldsymbol{\beta})\hat{R}(\mathbf{U}^{\mathrm{t}})\hat{S}(\boldsymbol{\Xi}_\kappa)\hat{D}(\boldsymbol{\alpha}_\kappa)$.
The first step to find its Bloch-Messiah decomposition is to find the Bogoliubov operators corresponding to the action of $\hat{O}_{\kappa}$ on the creation operator. In other words, we need to find $\mathbf{X}_{\kappa}, \mathbf{Y}_{\kappa}$ and $\mathbf{z}_{\kappa}$ such that,
\begin{align}
    \hat{\mathbf{a}}^{' \dagger} = \hat{O}_{\kappa}^{\dagger}\hat{\mathbf{a}}^{\dagger}\hat{O}_{\kappa} = \mathbf{X}_{\kappa}\hat{\mathbf{a}} + \mathbf{Y}_{\kappa}\hat{\mathbf{a}}^{\dagger} + \mathbf{z}_{\kappa} .
\end{align}
Then, by taking the singular value decomposition $\mathbf{X}_{\kappa} = \mathbf{V}_{\kappa} \mathrm{sinh}(\boldsymbol{\Sigma}_{\kappa}) \mathbf{W}_{\kappa}^\mathrm{t},\mathbf{Y}_{\kappa} = \mathbf{V}_{\kappa} \mathrm{cosh}(\boldsymbol{\Sigma}_{\kappa}) \mathbf{W}_{\kappa}^\dagger$, we will have,
\begin{align}
    \hat O_{\kappa} =  \hat{D}(\mathbf{z}_{\kappa})\hat{R}(\mathbf{V}_{\kappa})\hat S(\boldsymbol{\Sigma}_{\kappa}) \hat{R}(\mathbf{W}_{\kappa})^\dagger  ,
\end{align}
We remind here the actions of the multimode Gaussian operators on the ladder operators,
\begin{align}
    \hat S(\boldsymbol{\Xi})^{\dagger}\hat {\mathbf{a}}\hat S(\boldsymbol{\Xi}) &= \mathrm{cosh}(\boldsymbol{\Xi})\hat{\mathbf{a}} + \mathrm{sinh}(\boldsymbol{\Xi})\hat{\mathbf{a}}^{\dagger} , \\
    \hat D(\boldsymbol{\alpha})^{\dagger} \hat{\mathbf{a}} \hat D(\boldsymbol{\alpha}) &= \hat{\mathbf{a}} + \boldsymbol{\alpha} , \\ 
    \hat{R}(\mathbf{U})^{\dagger}\hat{\mathbf{a}}\hat{R}(\mathbf{U})&=\mathbf{U}\hat{\mathbf{a}} .
\end{align}
By applying these rules sequentially we get, 
\begin{align}
    \hat{\mathbf{a}}^{'\dagger} &= \hat O_\kappa ^{\dagger}\hat{\mathbf{a}}^\dagger \hat O_\kappa \nonumber \\
    &= (\hat R({\mathbf{U_2}})\hat S(\mathbf{\ln(L)}) \hat{R}(\mathbf{U}_1)\hat{D}(\boldsymbol{\beta})\hat{R}(\mathbf{U}^{\mathrm{t}})\hat{S}(\boldsymbol{\Xi}_\kappa)\hat{D}(\boldsymbol{\alpha}))^{\dagger} \left(\hat{\mathbf{a}}^{\dagger}\right)(\hat R({\mathbf{U_2}})\hat S(\mathbf{\ln(L)}) \hat{R}(\mathbf{U}_1)\hat{D}(\boldsymbol{\beta})\hat{R}(\mathbf{U}^{\mathrm{t}})\hat{S}(\boldsymbol{\Xi}_\kappa)\hat{D}(\boldsymbol{\alpha}))\nonumber \\
    &= (\hat S(\mathbf{\ln(L)}) \hat{R}(\mathbf{U}_1)\hat{D}(\boldsymbol{\beta})\hat{R}(\mathbf{U}^{\mathrm{t}})\hat{S}(\boldsymbol{\Xi}_\kappa)\hat{D}(\boldsymbol{\alpha}))^{\dagger} \left(\mathbf{U_2}\hat{\mathbf{a}}^{\dagger}\right)(\hat S(\mathbf{\ln(L)}) \hat{R}(\mathbf{U}_1)\hat{D}(\boldsymbol{\beta})\hat{R}(\mathbf{U}^{\mathrm{t}})\hat{S}(\boldsymbol{\Xi}_\kappa)\hat{D}(\boldsymbol{\alpha})) \nonumber  \\
    &= ( \hat{R}(\mathbf{U}_1)\hat{D}(\boldsymbol{\beta})\hat{R}(\mathbf{U}^{\mathrm{t}})\hat{S}(\boldsymbol{\Xi}_\kappa)\hat{D}(\boldsymbol{\alpha}))^{\dagger} \left(\mathbf{U_2}\mathrm{cosh}(\mathbf{\ln(L)})\hat{\mathbf{a}}^{\dagger}+\mathbf{U_2}\mathrm{sinh}(\mathbf{\ln(L)})\hat{\mathbf{a}}\right)( \hat{R}(\mathbf{U}_1)\hat{D}(\boldsymbol{\beta})\hat{R}(\mathbf{U}^{\mathrm{t}})\hat{S}(\boldsymbol{\Xi}_\kappa)\hat{D}(\boldsymbol{\alpha})) \nonumber \\
    &= ( \hat{D}(\boldsymbol{\beta})\hat{R}(\mathbf{U}^{\mathrm{t}})\hat{S}(\boldsymbol{\Xi}_\kappa)\hat{D}(\boldsymbol{\alpha}))^{\dagger} \left(\mathbf{U_2}\mathrm{cosh}(\mathbf{\ln(L)})\mathbf{U_1}\hat{\mathbf{a}}^{\dagger}+\mathbf{U_2}\mathrm{cosh}(\mathbf{\ln(L)})\mathbf{U_1}\hat{\mathbf{a}}\right)( \hat{D}(\boldsymbol{\beta})\hat{R}(\mathbf{U}^{\mathrm{t}})\hat{S}(\boldsymbol{\Xi}_\kappa)\hat{D}(\boldsymbol{\alpha})) \nonumber \\
    &= ( \hat{R}(\mathbf{U}^{\mathrm{t}})\hat{S}(\boldsymbol{\Xi}_\kappa)\hat{D}(\boldsymbol{\alpha}))^{\dagger} \left(\mathbf{U_2}\mathrm{cosh}(\mathbf{\ln(L)})\mathbf{U_1}\hat{\mathbf{a}}^{\dagger}+\mathbf{U_2}\mathrm{sinh}(\mathbf{\ln(L)})\mathbf{U_1}\hat{\mathbf{a}}+\mathbf{U_2}\exp(\mathbf{\ln(L)})\mathbf{U_1}\boldsymbol{\beta}\right)( \hat{R}(\mathbf{U}^{\mathrm{t}})\hat{S}(\boldsymbol{\Xi}_\kappa)\hat{D}(\boldsymbol{\alpha})) \nonumber  \\
    &= ( \hat{S}(\boldsymbol{\Xi}_\kappa)\hat{D}(\boldsymbol{\alpha}))^{\dagger} \left(\mathbf{U_2}\mathrm{cosh}(\mathbf{\ln(L)})\mathbf{U_1}\mathbf{U}^{\mathrm{t}}\hat{\mathbf{a}}^{\dagger}+\mathbf{U_2}\mathrm{sinh}(\mathbf{\ln(L)})\mathbf{U_1}\mathbf{U}^{\mathrm{t}}\hat{\mathbf{a}}+\mathbf{U_2}\mathbf{L}\mathbf{U_1}\boldsymbol{\beta}\right)( \hat{S}(\boldsymbol{\Xi}_\kappa)\hat{D}(\boldsymbol{\alpha})) \nonumber  \\
    &= \hat{D}(\boldsymbol{\alpha})^{\dagger} ([\mathbf{U_2}\mathrm{cosh}(\mathbf{\ln(L)})\mathbf{U_1}\mathbf{U}^{\mathrm{t}}\mathrm{cosh}(\boldsymbol{\Xi}_\kappa)+\mathbf{U_2}\mathrm{sinh}(\mathbf{\ln(L)})\mathbf{U_1}\mathbf{U}^{\mathrm{t}}\mathrm{sinh}(\boldsymbol{\Xi}_\kappa)]\hat{\mathbf{a}}^{\dagger}\nonumber  \\
    &+[\mathbf{U_2}\mathrm{sinh}(\mathbf{\ln(L)})\mathbf{U_1}\mathbf{U}^{\mathrm{t}}\mathrm{cosh}(\boldsymbol{\Xi}_\kappa)+\mathbf{U_2}\mathrm{cosh}(\mathbf{\ln(L)})\mathbf{U_1}\mathbf{U}^{\mathrm{t}}\mathrm{sinh}(\boldsymbol{\Xi}_\kappa)]\hat{\mathbf{a}}+\mathbf{U_2}\mathbf{L}\mathbf{U_1}\boldsymbol{\beta})\hat{D}(\boldsymbol{\alpha}) \nonumber  \\
    &= \left[\mathbf{U_2}\mathrm{cosh}(\mathbf{\ln(L)})\mathbf{U_1}\mathbf{U}^{\mathrm{t}}\mathrm{cosh}(\boldsymbol{\Xi}_\kappa)+\mathbf{U_2}\mathrm{sinh}(\mathbf{\ln(L)})\mathbf{U_1}\mathbf{U}^{\mathrm{t}}\mathrm{sinh}(\boldsymbol{\Xi}_\kappa)\right]\hat{\mathbf{a}}^{\dagger}\nonumber  \\
    &+\left[\mathbf{U_2}\mathrm{sinh}(\mathbf{\ln(L)})\mathbf{U_1}\mathbf{U}^{\mathrm{t}}\mathrm{cosh}(\boldsymbol{\Xi}_\kappa)+\mathbf{U_2}\mathrm{cosh}(\mathbf{\ln(L)})\mathbf{U_1}\mathbf{U}^{\mathrm{t}}\mathrm{sinh}(\boldsymbol{\Xi}_\kappa)\right]\hat{\mathbf{a}} \nonumber \\
    &+ [\mathbf{U_2}\mathrm{cosh}(\mathbf{\ln(L)})\mathbf{U_1}\mathbf{U}^{\mathrm{t}}\mathrm{cosh}(\boldsymbol{\Xi}_\kappa)+\mathbf{U_2}\mathrm{sinh}(\mathbf{\ln(L)})\mathbf{U_1}\mathbf{U}^{\mathrm{t}}\mathrm{sinh}(\boldsymbol{\Xi}_\kappa) \nonumber \\
    &+\mathbf{U_2}\mathrm{sinh}(\mathbf{\ln(L)})\mathbf{U_1}\mathbf{U}^{\mathrm{t}}\mathrm{cosh}(\boldsymbol{\Xi}_\kappa)+\mathbf{U_2}\mathrm{cosh}(\mathbf{\ln(L)})\mathbf{U_1}\mathbf{U}^{\mathrm{t}}\mathrm{sinh}(\boldsymbol{\Xi}_\kappa)]\boldsymbol{\alpha}+\mathbf{U_2}\mathbf{L}\mathbf{U_1}\boldsymbol{\beta} \nonumber \\
    &= \mathbf{X}_\kappa\hat{\mathbf{a}} + \mathbf{Y}_\kappa\hat{\mathbf{a}}^{\dagger} + \mathbf{z}_\kappa ,
\end{align}
where,
\begin{align}
    \mathbf{X}_\kappa &= \mathbf{U_2}\mathrm{sinh}(\mathbf{\ln(L)})\mathbf{U_1}\mathbf{U}^{\mathrm{t}}\mathrm{cosh}(\boldsymbol{\Xi}_\kappa)+\mathbf{U_2}\mathrm{cosh}(\mathbf{\ln(L)})\mathbf{U_1}\mathbf{U}^{\mathrm{t}}\mathrm{sinh}(\boldsymbol{\Xi}_\kappa) , \\ 
    \mathbf{Y}_\kappa &= \mathbf{U_2}\mathrm{cosh}(\mathbf{\ln(L)})\mathbf{U_1}\mathbf{U}^{\mathrm{t}}\mathrm{cosh}(\boldsymbol{\Xi}_\kappa)+\mathbf{U_2}\mathrm{sinh}(\mathbf{\ln(L)})\mathbf{U_1}\mathbf{U}^{\mathrm{t}}\mathrm{sinh}(\boldsymbol{\Xi}_\kappa) , \\
    \mathbf{z}_\kappa &= \mathbf{U_2}\mathbf{L}\mathbf{U_1}\mathbf{U}^{\mathrm{t}}\exp(\boldsymbol{\Xi}_\kappa)\boldsymbol{\alpha}_\kappa + \mathbf{U_2}\mathbf{L}\mathbf{U_1}\boldsymbol{\beta} ,
\end{align}
as $\mathbf{\boldsymbol{\alpha}}_\kappa$ and $\mathbf{\boldsymbol{\beta}}$ are real vectors and $\mathrm{cosh}(\mathbf{A})+\mathrm{sinh}(\mathbf{A}) = \exp(\mathbf{A})$ for any matrix $\mathbf{A}$. 
Finally, we have,
\begin{align}
    \hat O_\kappa &= \hat{D}(\mathbf{z}_\kappa)\hat{R}(\mathbf{V}_\kappa)  \hat S(\boldsymbol{\Sigma}_\kappa) \hat{R}(\mathbf{W}_\kappa)^\dagger \nonumber \\
    &= \hat{R}(\mathbf{V}_\kappa)  \hat S(\boldsymbol{\Sigma}_\kappa) \hat{R}(\mathbf{W}_\kappa)^\dagger \hat D(\boldsymbol{\gamma}_\kappa) ,
\end{align}
where $\boldsymbol{\gamma}_\kappa = \mathbf{W}_\kappa\exp(\boldsymbol{-\Sigma}_\kappa)\mathbf{V}_{\kappa}^\dagger\mathbf{z}_\kappa$, which leads to,
\begin{align}
    f_{\mathbf{m^{'}}}(\kappa) = \exp(2\kappa \mu^{(0)})\prod^M_{j=1} |C_{j,\kappa}|^2 \left|\bra{\mathbf{m^{'}}}\hat{R}(\mathbf{V}_\kappa)  \hat S(\boldsymbol{\Sigma}_\kappa) \hat{R}(\mathbf{W}_{\kappa})^\dagger \hat{D}(\boldsymbol{\gamma}_\kappa)\ket{\mathbf{0}}\right|^2 .
\end{align}
\section{\label{app:appedix Benzene_e2g_para} Molecular parameters for the linear Herzberg-Teller simulation of benzene}
The molecular parameters for the linear Herzberg-Teller simulation of benzene are given as  following, 
{\scriptsize
\begin{align}
    \mathbf{U_\mathrm{D}} &= \begin{pmatrix}
     -8.6323\times10^{-6}&9.9999\times10^{-1}&1.7996\times10^{-5}&-1.2862\times10^{-3}&4.5475\times10^{-6}&
                -8.8874\times10^{-4}&-1.9329\times10^{-3}&-3.1563\times10^{-6} \\
              9.9999\times10^{-1}&4.6727\times10^{-6}&1.2715\times10^{-3}&-8.2931\times10^{-6}&8.8318\times10^{-4}&
               -1.7281\times10^{-6}&-2.0265\times10^{-5}&1.9217\times10^{-3} \\
              -1.2114\times10^{-3}&-2.2308\times10^{-5}&9.9881\times10^{-1}&1.2296\times10^{-5}&-4.8525\times10^{-2}&
                2.2123\times10^{-5}&1.7170\times10^{-5}&-2.9745\times10^{-3} \\
              1.0355\times10^{-5}&1.2122\times10^{-3}&-1.0929\times10^{-6}&9.9881\times10^{-1}&9.1339\times10^{-6}&
                -4.8506\times10^{-2}&-2.9613\times10^{-3}&8.6992\times10^{-7} \\
              -9.4803\times10^{-4}&-6.0921\times10^{-6}&4.8519\times10^{-2}&9.1307\times10^{-6}&9.9877\times10^{-1}&
                -1.3846\times10^{-4}&-4.0148\times10^{-6}&9.2288\times10^{-3} \\
              -6.0924\times10^{-6}&9.6095\times10^{-4}&2.2275\times10^{-6}&4.8555\times10^{-2}&1.4398\times10^{-4}&
                9.9877\times10^{-1}&9.2157\times10^{-3}&-1.5228\times10^{-5} \\
              -1.9385\times10^{-3}&-5.1031\times10^{-6}&2.4998\times10^{-3}&7.1391\times10^{-6}&-9.3712\times10^{-3}&
                3.3886\times10^{-7}&-2.4854\times10^{-6}&9.9995\times10^{-1} \\
              -1.1184\times10^{-5}&1.9114\times10^{-3}&4.0060\times10^{-6}&2.5096\times10^{-3}&1.7972\times10^{-5}&
                -9.3588\times10^{-3}&9.9995\times10^{-1}&1.0993\times10^{-5}
 \end{pmatrix} ,
 \end{align}}
{\footnotesize
 \begin{align}
    \boldsymbol{\delta} &= (7.4613\times10^{-6},-3.1429\times10^{-5},-3.7674\times10^{-5},3.0782\times10^{-5},2.3498\times10^{-5},
-3.4351\times10^{-5},2.8415\times10^{-5},-6.8897\times10^{-6})^{\mathrm{t}},
\end{align}}
\begin{align}
    \boldsymbol{\omega} &= (3369.2220,3369.2220,1730.1965,1730.1964,1263.3599,1263.3599,646.1398,646.1398)^{\mathrm{t}} , \\
    \boldsymbol{\omega^{'}} &= (3389.0368,3389.0368,1665.3003,1665.3003,1236.7103,1236.7103,575.1367,575.1367)^{\mathrm{t}} ,\\
    \boldsymbol{\hat{\mu}}(\mathbf{\hat{Q}}) &= (\hat{\mu}_{x}(\mathbf{\hat Q}),\hat{\mu}_{y}(\mathbf{\hat Q}),0)^{\mathrm{t}} = (\mu_{x}^{(0)} + \boldsymbol{\mu^{(1)}_{x}}\cdot\mathbf{\hat{Q}},\mu_{y}^{(0)} + \boldsymbol{\mu^{(1)}_{y}}\cdot\mathbf{\hat{Q}}, 0)^{\mathrm{t}},
\end{align}
with $\mu_{x}^{(0)} = 0.0000 \: \mathrm{D}, \; \mu_{y}^{(0)} = 0.0000 \; \mathrm{D}$  and $\boldsymbol{\mu_{x}^{(1)}} = (0.3054,0.0000,0.0000,0.1795,0.1190,0.0000,0.0000,0.5710)^{\mathrm{t}}, \; \boldsymbol{\mu_{y}^{(1)}} = (0.0000,0.3054,-0.1795,0.0000,0.0000,-0.1190,-0.5710,0.0000)^{\mathrm{t}}$ expressed in $\mathrm{D/(u^{\tfrac{1}{2}}\angstrom)}$.

\end{widetext}

%

\end{document}